\documentclass{ws-ijmpe}
\usepackage{hyperref}
\usepackage[super,compress]{cite}
\usepackage{multicol}
\usepackage{multirow}
\begin{document}

\markboth{Manuel Pav\'on Valderrama}{Power Counting and Renormalization}

\catchline{}{}{}{}{}

\title{Power Counting and Wilsonian Renormalization \\
in Nuclear Effective Field Theory}

\author{Manuel Pav\'on Valderrama}

\address{Institut de Physique Nucl\'eaire, CNRS-IN2P3, Univ. Paris-Sud, \\
Universit\'e Paris-Saclay, F-91406 Orsay Cedex, France\\
School of Physics and Nuclear Energy Engineering, \\
Beihang University, Beijing 100191, China\\
pavonvalderrama@ipno.in2p3.fr}

\maketitle

\begin{history}
\received{Day Month Year}
\revised{Day Month Year}
\end{history}

\begin{abstract}
Effective field theories are the most general tool
for the description of low energy phenomena.
They are universal and systematic:
they can be formulated for any low energy systems we can think of
and offer a clear guide on how to calculate predictions
with reliable error estimates,
a feature that is called power counting.
These properties can be easily understood in Wilsonian renormalization,
in which effective field theories are the low energy renormalization
group evolution of a more fundamental ---perhaps unknown or unsolvable---
high energy theory.
In nuclear physics they provide the possibility of a theoretically sound
derivation of nuclear forces without having to solve
quantum chromodynamics explicitly.
However there is the problem of how to organize calculations
within nuclear effective field theory: the traditional knowledge
about power counting is perturbative but nuclear physics is not.
Yet power counting can be derived in Wilsonian renormalization
and there is already a fairly good understanding of
how to apply these ideas to non-perturbative phenomena and
in particular to nuclear physics.
Here we review a few of these ideas, explain power counting
in two-nucleon scattering and reactions with external probes
and hint at how to extend the present analysis
beyond the two-body problem.
\end{abstract}


\ccode{PACS numbers:}


\section{Introduction}

The structure, properties and reactions of nuclei and nuclear matter
depend on the dynamics of the nucleons~\cite{Machleidt:2001rw}.
This is the reason why the derivation of the nuclear forces
is probably the most important problem of nuclear physics.
After the discovery of quantum chromodynamics (QCD)
--- the fundamental theory of strong interactions ---
a solid theoretical understanding of the nuclear force
should be grounded on QCD, either directly or indirectly.
Lattice QCD represents the direct, computational derivation:
the interaction of quarks and gluons is not analytically solvable
at the distances that are characteristic for nuclear physics
but it is numerically solvable at the expense of huge computational resources.
Recent progress in this front
is exciting~\cite{Beane:2011iw,Beane:2012vq,Beane:2013br,Aoki:2012tk,Yamazaki:2015asa}.
The indirect derivation requires to explain the nuclear interaction
without explicitly solving QCD.
Yet QCD must enter indirectly in the picture. Otherwise we will end up
with a phenomenological description instead of a theoretical explanation.

Physics as a science depends to a great extent
on the existence of scale separation in nature.
One can describe the properties of atoms without explicit knowledge of
the composite nature and internal structure of the nuclei within.
The nucleus is indeed much smaller than the atom containing it, i.e.
there is an excellent separation of scales.
Analogously, one can describe the dynamics of nucleons and pions
without knowing the details of the strong interaction of
the quarks and gluons inside them.
However the average distance of nucleons in a nucleus
--- about $1-2\,{\rm fm}$ --- is not that different from
the size of the nucleon or the wavelength of the quarks
and gluons inside, maybe $0.5\,{\rm fm}$.
Without a clear separation of scales the development of
satisfactory theoretical explanations to physical phenomena
becomes more difficult.
As a consequence the description of nuclei is
less clear and more involved than that of atoms.

Effective field theories (EFTs) are the standard theoretical tool
to exploit the separation of scales of a physical system with
the intention of building the most general description
of it at low energies~\cite{Beane:2000fx,Bedaque:2002mn,Machleidt:2011zz}.
If we call the low energy scale $Q$ and the high energy scale $M$,
an EFT provides a power expansion in terms of $Q/M$ of
all the physical quantities of a system.
For that one considers first all the possible interaction terms
in the Lagrangian that are compatible with
the low energy symmetries of the system.
Then one orders the infinite Feynman diagrams obtained in the previous step
according to their expected size. The method by which we estimate
the size of the diagrams is called power counting.
While writing the diagrams is trivial, their power counting is not.

The connection of the EFT to the fundamental theory at the scale $M$
is provided by renormalization, the core idea of EFT.
In its standard formulation renormalization deals with ultraviolet (UV)
divergences in the Feynman diagrams of the EFT.
To remove the divergences one includes an UV cut-off and allows the couplings
in the Lagrangian to depend on the cut-off.
If the calculation of the observable quantities of the EFT is independent
of the cut-off then the EFT is renormalizable.
Power counting is decided according to how we have to arrange
calculations to remove the divergences at each order
in the expansion.
Wilsonian renormalization~\cite{Wilson:1973jj}
provides an alternative but equivalent formulation.
Here the starting point
is the independence of observables with respect to the cut-off.
In this case it is the calculation of the couplings under the assumption of
cut-off independence that leads to the size of these couplings
at low energies and to their power counting~\cite{Polchinski:1983gv,Birse:1998dk,Barford:2002je,Birse:2005um}.
This is referred to as renormalization group: the focus is
on the evolution of the couplings as the cut-off changes,
not on the divergences.
In Wilsonian renormalization the cut-off runs from the high to
the low energy scale, from $M$ to $Q$.
This is counterintuitive from the standard point of view,
where the cut-off runs from $Q$ to $M$ with the purpose of
finding out whether there are UV divergences.
Yet they are equivalent. The cut-off can either run to the ultraviolet or
the infrared (IR). As far as the observables are independent of the cut-off 
we end up with identical power countings.
The starting point in Wilsonian renormalization can be either an EFT or
the fundamental theory.
The advantage in the first case is that power counting can be determined
without a complete order-by-order calculation of observables.
In the second case there is the possibility of evolving a fundamental theory
from $M$ to $Q$, which amounts to uncovering the EFT by means of
a concrete calculation.
Of course this is only possible in the few cases where the fundamental theory
is known or easily solvable 
(a nice example can be found in Ref.~\refcite{Polchinski:1983gv}).
This manuscript is dedicated to Wilsonian renormalization
in nuclear EFT~\cite{Birse:1998dk,Barford:2002je,Birse:2005um,PavonValderrama:2007nu,Valderrama:2014vra}:
even though it is less well-known than the standard idea of
removing divergences, it can provide a clearer interpretation
of power counting and the role of the cut-off in EFT.
 
In nuclear physics the EFT usually contains nucleon and pion fields~\footnote{
There are also nuclear EFTs with additional fields
-- such as the delta isobar -- or without the pion field (the pionless EFT).}
that are constrained by chiral symmetry, a low energy symmetry
of QCD that is exact in the limit of massless
$u$, $d$ and sometimes $s$ quarks.
The problem is that historically renormalization has been only well understood
in the case of systems that are perturbative~\cite{Weinberg:1978kz,Polchinski:1983gv}.
This is the case in hadron physics for processes involving at most one baryon,
where chiral perturbation theory (ChPT)~\cite{Bernard:1995dp}, the standard
EFT for low energy hadronic processes, is used.
But in nuclear physics the existence of the deuteron and the virtual state
(the $^1S_0$ singlet), not to mention the few thousand known nuclei, indicates
that the nuclear force is non-perturbative.
Besides there is the additional problem that EFT entails nuclear forces
that are strongly divergent at short distances.
Thus it is not a surprise that progress in nuclear EFT has been
full of unexpected turns and controversies.
Recently we have begun to have a solid grasp of
the non-perturbative renormalization of the EFT potentials~\cite{Beane:2000wh,PavonValderrama:2005gu,PavonValderrama:2005wv,PavonValderrama:2005uj,Long:2007vp}
and how to organize the power counting in this situation~\cite{Nogga:2005hy,Birse:2005um,Valderrama:2009ei,Valderrama:2011mv,Long:2011qx,Long:2011xw,Long:2012ve},
but even these advances have been the subject of
debate~\cite{Epelbaum:2006pt,Epelbaum:2009sd,Machleidt:2010kb,Marji:2013uia}.
Here we will review power counting from the perspective of
the renormalization group.

Historically events have unfolded in a zig-zag pattern.
Weinberg made the first proposal
for a nuclear EFT~\cite{Weinberg:1990rz,Weinberg:1991um},
which includes the iteration of the EFT potential (at least at lowest order).
This serves to capture the non-perturbative character of nuclear interactions
but in exchange requires non-perturbative renormalization.
As previously said, this has been the source of a few surprises.
Kaplan, Savage and Wise (KSW) discovered a subtle but nonetheless
serious inconsistency with the Weinberg proposal~\cite{Kaplan:1996xu}.
These authors also developed a new formulation of nuclear EFT,
the KSW counting~\cite{Kaplan:1998tg,Kaplan:1998we},
which is free from that inconsistency.
However the convergence of the KSW counting in the triplet partial waves
happened to be unsatisfying to say the least~\cite{Fleming:1999ee}.
The community turned back to the Weinberg proposal in search
for phenomenological success~\cite{Entem:2003ft,Epelbaum:2004fk}.
But later Nogga, Timmermans and van Kolck~\cite{Nogga:2005hy}
discovered that the Weinberg proposal contains a new,
more conspicuous inconsistency at the lowest order:
it is not renormalizable in some P- and D-waves~\footnote{Actually,
the KSW inconsistency already indicated that the Weinberg proposal
is not renormalizable at the lowest order. However this problem
does not directly affect two-nucleon scattering.}.
New developments about renormalizability followed~\cite{Birse:2005um,PavonValderrama:2005gu,PavonValderrama:2005wv,PavonValderrama:2005uj,Long:2007vp}
that made finally possible a consistent nuclear EFT
with good convergence properties~\cite{Valderrama:2009ei,Valderrama:2011mv,Long:2011qx,Long:2011xw,Long:2012ve}.
Despite these advancements, there is an ongoing debate about whether
these problems are relevant and whether it would be simply more sensible
to reinterpret renormalizability for non-perturbative
problems in a different way~\cite{Epelbaum:2006pt,Epelbaum:2009sd,Machleidt:2010kb,Marji:2013uia}.
We will not discuss these new developments, except for a brief comment.
Here we are mostly concerned about the derivation of EFT power counting
from a specific set of renormalization tools, which happen to be
more than enough to make nuclear EFT work at the theoretical level.
From this perspective the previous ideas, though interesting,
do not appear to be totally necessary.

This manuscript is organized as follows:
in Sect. II we introduce Wilsonian renormalization for the particular case
of non-relativistic scattering of two particles.
Part of it is general and part of it is specific to nuclear physics.
We also discuss the relationship of power counting with the anomalous dimension
of couplings and the relationship between Wilsonian renormalization and
the more standard approach of removing ultraviolet divergences.
In Sect. III we extend the results beyond the two-nucleon system, in particular
to the deuteron electroweak reactions and to the three-body problem.
Finally we summarize our conclusions. We also include an appendix discussing
the derivation of a particular equation in this manuscript.

\section{Wilsonian Renormalization}

Here we illustrate how Wilsonian renormalization works 
for non-relativistic s-wave scattering~\cite{Birse:1998dk,Barford:2002je,Birse:2005um,PavonValderrama:2007nu}.
The starting point is a ``fundamental theory''.
For a non-relativistic two-body system the equivalent of a fundamental theory
is the non-relativistic potential $V$.
To obtain the scattering amplitudes we solve the Schr\"odinger equation
at finite momentum $k$
\begin{eqnarray}
-u_k^{''}(r) + 2\mu\,V(r)\,u_k(r) = k^2 u_k(r) \, ,
\label{eq:schro-fundamental}
\end{eqnarray}
where $u_k$ is the reduced wave function, $\mu$ the reduced mass,
$k$ the center of mass momentum and $V(r)$ the underlying potential,
which we assume to be known at all distances.
As we are considering s-wave scattering there is no centrifugal term.
We solve this equation with the regular boundary condition at the origin
\begin{eqnarray}
u(0) = 0 \, .
\end{eqnarray}
Finally the phase shift can be extracted from the asymptotic wave function
\begin{eqnarray}
u_k(r) \to \sin{(k r + \delta)} \quad \mbox{for $r \to \infty$.}
\end{eqnarray}

Wilsonian renormalization works as follows.
In a first step we include a cut-off $r_c$ as a separation scale
\begin{eqnarray}
V(r) \to V(r; r_c) = V(r)\,\theta(r - r_c) \, .
\end{eqnarray}
We will consider that the physics at distances shorter
than the cut-off $r < r_c$ is unknown.
Of course if we cut the potential for $r < r_c$
the physical observables will change.
We want to prevent this from happening. 
In a second step we include a new piece in the potential that counteracts
the loss of information from having a cut-off and 
keeps the observables unchanged.
This extra piece is the contact-range potential,
which can take many parametrizations.
For simplicity we choose the following form for the contacts
\begin{eqnarray}
V_C(r; r_c) = \frac{\delta(r-r_c)}{4 \pi r_c^2}\,\sum_{n = 0}^{\infty} C_{2n}(r_c)
\,k^{2n} \, ,
\end{eqnarray}
that is, an energy-dependent delta shell potential.
Now we solve the Schr\"odinger equation with the ``renormalized'' potential
\begin{eqnarray}
V_R(r; r_c) = V(r; r_c) + V_C(r; r_c) \, .
\end{eqnarray}
For distances below $r_c$ we have a free Schr\"odinger equation
\begin{eqnarray}
-u_k^{''}(r) = k^2 u_k(r) \, ,
\end{eqnarray}
with the regular solution
\begin{eqnarray}
u_k(r) = \sin{(k r)} \, .
\end{eqnarray}
For distances above $r_c$ we have the original Schr\"odinger equation, i.e.
Eq.~(\ref{eq:schro-fundamental}).
Finally at $r = r_c$ the delta-shell potential $V_C$ generates a discontinuity
in the first derivative of the wave function that takes the form
\begin{eqnarray}
\frac{u_k'(r_c^{+})}{u_k(r_c^{+})} - \frac{u_k'(r_c^{-})}{u_k(r_c^{-})} =
\frac{\mu}{2\pi r_c^2}\,\sum_{n} C_{2n}(r_c) k^{2n} \, , \label{eq:RGE}
\end{eqnarray}
where $r_c^{\pm}$ refers to $r_c \pm \epsilon$, with $\epsilon \to 0$.
A derivation can be found in \ref{app:delta}.
This is the renormalization group equation (RGE)
for the contact-range coupling $C_{2n}$.
The RGE we have written above is exact: the starting point is
the full potential $V$ and we want to check what type of
contact interaction we have to include to account
for the existence of a cut-off radius $r_c$.
For $r_c = 0$ we have the boundary condition $C_{2n}(0) = 0$: we know
the potential at all distances and there is no need
for the contact-range couplings.
As we increase the cut-off radius, we will need non-vanishing $C_{2n}(r_c)$
couplings to account for the missing physics.

The reason we are interested in Wilsonian renormalization is because
we want to know how physics looks like at large distances in general.
We want to describe phenomena at low energies
regardless of which is the fundamental theory at high energies.
With Wilsonian renormalization we can build a theory for distances larger
than the cut-off ($r \geq r_c$) that is equivalent
to the fundamental theory for momenta $k\,r_c < 1$.
In this context it is useful to define the soft and hard scales $Q$ and $M$.
The soft scale $Q$ is the characteristic momentum of the low energy physics
we want to describe, while the hard scale $M$ is the natural momentum scale
of the fundamental theory.
If we solve the RGE for $Q \, r_c \to 1$ we will be able to derive the 
kind of generic low energy theory we are interested in.

To solve the RGE and obtain the contact range couplings
we simply have to make an ansatz for the wave function $u_k$.
The simplest case is provided by a theory in which the underlying
potential has a finite range set by the hard scale $M$
\begin{eqnarray}
V(r) \to 0 \quad \mbox{for $M r \gg 1$.}
\end{eqnarray}
With this is mind we see that the wave functions for $M r \gg 1$ are given by
\begin{eqnarray}
u_k(r) \to \sin{(k r + \delta)} \, ,
\end{eqnarray}
where $\delta$ is the phase shift of the fundamental potential $V$.
Therefore the solution of the RGE for $M r_c \gg 1$ is
\begin{eqnarray}
k\,\cot{(k r_c + \delta)} - k\,{\cot{k r_c}} =
\frac{\mu}{2\pi r_c^2}\,\sum_{n} C_{2n}(r_c) k^2 \, .
\end{eqnarray}
For finding the running of the individual $C_{2n}(r_c)$ couplings
we expand the RGE in powers of $k^2$.
We first take into account that $V$ is a finite-range potential
and the effective range expansion applies (for $k < M$)
\begin{eqnarray}
k\,\cot{\delta} = -\frac{1}{a_0} + \frac{1}{2}\,r_0\,k^2 +
\sum_{n=2}^{\infty} v_n \, k^{2n} \, ,
\end{eqnarray}
where $a_0$ is the scattering length, $r_0$ the effective range and
$v_n$ the shape coefficients.
Now we expand and get the set of equations
\begin{eqnarray}
\frac{1}{r_c - a_0} - \frac{1}{r_c} &=& \frac{\mu}{2\pi r_c^2}\,C_0(r_c) \, , \\
\frac{a_0^2\,\frac{r_0}{2}}{(r_c - a_0)^2} +
\frac{P_2(r_c, a_0)}{(r_c - a_0)^2} + 
\frac{r_c}{3}
&=& \frac{\mu}{2\pi r_c^2}\,C_2(r_c) \, , \\
\frac{a_0^2\,v_2}{(r_c - a_0)^2} + 
\frac{P_4(r_c, a_0, r_0)}{(r_c - a_0)^3}
+ 
\frac{r_c^3}{45}
&=& \frac{\mu}{2\pi r_c^2}\,C_4(r_c) \, ,
\end{eqnarray}
plus analogous equations for the higher order couplings.
In the equations above $P_2(r_c, a_0)$ and $P_4(r_c, a_0, r_0)$ are polynomials
of the cut-off $r_c$ and the effective range coefficients.
They are easy to calculate but they are not included here because
their exact form is inconsequential for the analysis that follows.

The previous equations are generic solutions for an arbitrary finite range
potential $V$.
However the counting of the couplings $C_{2n}(r_c)$ as $Q r_c \to 1$
depends on which is the size of the effective range coefficients
and in particular the scattering length.
In general the size of the effective range coefficients is known to scale
according to the range of the potential (therefore the name):
\begin{eqnarray}
M r_0 \sim 1  \quad \mbox{and} \quad M^{2n+1} v_n \sim 1 \, .
\end{eqnarray}
The exception is the scattering length $a_0$ that can take any value,
more so if there is non-perturbative physics.
Thus we distinguish two possibilities: 
\begin{eqnarray}
M a_0 \sim 1 \quad \mbox{or} \quad Q a_0 \sim 1 \, .
\end{eqnarray}
The first one is a scattering length of natural size and
the second an unnaturally large scattering length, which is
what happens for instance if there is a bound state near the threshold.

If the scattering length is of order $1/M$ we are entitled to expand
in powers of $a_0 / r_c$ because $M r_c \gg 1$.
We obtain
\begin{eqnarray}
C_0(r_c) = \frac{2\pi}{\mu}\,a_0\,\left[ 1 + \frac{a_0}{r_c} +
\mathcal{O}\left( \frac{a_0^2}{r_c^2}\right) \right] \, ,
\end{eqnarray}
that is, $C_0(r_c)$ scales as $1/M^2$.
If we analyze now the subleading couplings $C_{2n}(r_c)$, we find for $C_2$
\begin{eqnarray}
C_2(r_c) = \frac{2\pi}{\mu}\,a_0^2\,\frac{r_0}{2}\,\left[ 1
+ 2\,\frac{a_0}{r_c} + \mathcal{O}\left( \frac{a_0^2}{r_c^2}\right)
\right] + C_2^R(r_c; a_0) \, ,
\end{eqnarray}
that is, $C_{2}$ scales as $1/M^4$.
In the equation above $C_2^R$ is a ``redundant'' piece of the coupling $C_2$
that does not contain information about a the effective range $r_0$.
The function of $C_2^R$ is to absorb the cut-off dependence
that the $C_0$ coupling generates at finite energy.
The $C_2^R$ piece of $C_2$ is inessential for power counting.
For $C_{4}$ we have
\begin{eqnarray}
C_4(r_c) = \frac{2\pi}{\mu}\,a_0^2\,v_2\,\left[ 1
+ 2\,\frac{a_0}{r_c} + \mathcal{O}\left( \frac{a_0^2}{r_c^2}\right)
\right] + C_4^R(r_c; a_0, r_0) \, ,
\end{eqnarray}
which scales as $1/M^6$ and where $C_4^R$ is analogous to $C_2^R$, only
that it absorbs the residual cut-off dependence of $C_0$ and $C_2$.
For the higher order couplings we have $C_{2n} \sim 1/M^{2n+2}$.

The other possibility is that the scattering length is large: $Q a_0 \sim 1$.
Now the cut-off and the scattering length can have the same size and
we are not allowed to expand in powers of $a_0 / r_c$.
If we solve the RGE for $C_0$ we obtain
\begin{eqnarray}
C_0(r_c) = \frac{2\pi}{\mu}\,\frac{r_c\,a_0}{r_c - a_0} \, ,
\end{eqnarray}
which means $C_0 \sim 1/(M Q)$, an enhancement of one power of $M/Q$.
For the $C_{2}$ coupling we get
\begin{eqnarray}
C_{2}(r_c) &=& \frac{2\pi}{\mu}\,\,a_0^2\,\frac{r_0}{2}\,
\frac{r_c^2}{(r_c - a_0)^2} + C_2^R(r_c; a_0) \, ,
\end{eqnarray}
which entails $C_{2} \sim 1/(M^{2} Q^2)$,
an enhancement of two powers of $M/Q$.
For the $C_4$ coupling we find
\begin{eqnarray}
C_{4}(r_c) &=& \frac{2\pi}{\mu}\,\,a_0^2\,v_2\,
\frac{r_c^2}{(r_c - a_0)^2} + C_4^R(r_c; a_0, r_0) \, ,
\end{eqnarray}
and in general for the $C_{2n}$ we have 
\begin{eqnarray}
C_{2n}(r_c) &=& \frac{2\pi}{\mu}\,\,a_0^2\,v_{n}\,
\frac{r_c^2}{(r_c - a_0)^2} + C_{2n}^R(r_c; a_0, r_0, \dots , v_{n-1}) \, ,
\end{eqnarray}
which implies a $M^2 / Q^2$ enhancement over the natural case.

The first implementation of this type of RG analysis of the couplings
in nuclear EFT is due to Birse, McGovern and Richardson~\cite{Birse:1998dk},
who formulated the RGEs in momentum space.
Instead of imposing the invariance of the phase shifts with respect to the
regulator, their analysis requires the invariance of
the full off-shell T-matrix.
For contact-range interactions both conditions are equivalent:
on-shell renormalization implies off-shell renormalization.
Probably this is the case too for finite-range potentials
(it has been proved for potentials that have
power-law divergences near the origin~\cite{Entem:2009mf}).
The analysis of the RGEs in momentum space is pretty convoluted though.
The analysis of the coordinate space RGEs of
Ref.~\refcite{PavonValderrama:2007nu} is simpler
as it only depends on the Schr\"odinger equation and its solutions.
It connects the RGEs with the cut-off dependence of the observables
after an arbitrary number of contact-range operators are included.
But at the same time it neglects how the RGEs relate
to the power counting of the couplings.
The purpose of this section has been to close this gap
and to translate the RG analysis of Ref.~\refcite{Birse:1998dk}
from momentum to coordinate space,
attempting to make its interpretation clearer in the way.

\subsection{Low Energy Effective Field Theory and Power Counting}

The question we wanted to answer is:
what kind of low energy theory does one derive from the RGEs?
The answer involves two ingredients.
The first is a non-relativistic potential for the low energy theory,
the effective potential:
\begin{eqnarray}
V_{\rm EFT}(r; r_c) = V_C(r; r_c) = \frac{\delta(r - r_c)}{4\pi r_c^2}\,
\sum_{n=0}^{\infty} C_{2n}(r_c) k^{2n} \, ,
\end{eqnarray}
that is, the contact-range potential that compensates the cut-off dependence.
The fundamental potential $V$ does not enter into the effective potential
for the simple reason that it vanishes at large distances ($M r \gg 1$).
The second ingredient is the size of the couplings, that we have already
calculated from the RGEs.
As a consequence of the scaling properties of the couplings
we can write the effective potential as a power series
in $Q/M$.
Let us consider the example of a theory with a natural scattering length,
for which we have
\begin{eqnarray}
C_{2n}(r_c) \sim \frac{1}{M^{2n+2}} \, .
\end{eqnarray}
If we take into account the typical factors of $\pi$ and the reduced mass
that are common in non-relativistic scattering,
the previous scaling allows to rewrite
the $C_{2n}$ couplings as 
\begin{eqnarray}
C_{2n}(r_c) = \frac{2 \pi}{\mu}\,\frac{c_{2n}(r_c)}{M^{2n+1}} \, ,
\end{eqnarray}
where $c_{2n}(r_c)$ is a number of $\mathcal{O}(1)$.
Now we plug this expression into the effective potential. We arrive to
\begin{eqnarray}
V_{\rm EFT}(r; r_c) = \frac{\delta(r - r_c)}{4\pi r_c^2}\,
\frac{2\pi}{\mu\,M}\,\sum_{n=0}^{\infty} c_{2n}(r_c)\,
{\left( \frac{Q}{M} \right)}^{2n} \, ,
\end{eqnarray}
that is, a power series in $Q/M$.
For large scattering length we have instead
\begin{eqnarray}
C_0(r_c) &=& \frac{2 \pi}{\mu Q}\,c_0(r_c) \, , \\
C_{2n}(r_c) &=& \frac{2 \pi}{\mu Q^2}\,\frac{c_{2n}(r_c)}{M^{2n-1}} 
\quad \mbox{for $n \geq 1$}\, ,
\end{eqnarray}
with $c_0(r_c) = \mathcal{O}(1)$ and
$c_{2n}(r_c) = \mathcal{O}(1)$ for $Q r_c \sim 1$.
In this case the expansion of the potential reads
\begin{eqnarray}
V_{\rm EFT}(r; r_c) = \frac{\delta(r - r_c)}{4\pi r_c^2}\,
\frac{2\pi}{\mu\,Q}\,\left[
c_0(r_c) + 
\sum_{n=1}^{\infty} c_{2n}(r_c)\,
{\left( \frac{Q}{M} \right)}^{2n-1} \right] \, .
\end{eqnarray}
Independently of the power counting of the couplings
we end up with a series that converges for $Q < M$.

This idea of arranging the effective potential as a power series
extends to every physical quantity we can think of.
It is a fundamental concept in EFT, the reason why calculations
are systematic.
We can predict observable quantities up to a given degree of accuracy,
that is, up to a given power of the expansion parameter $Q / M$.
The calculations are organized as to only include the operators
that contribute within the accuracy goals we have set up
in the first place.
We can illustrate this concept with the phase shift.
In perturbation theory the phase shift is expanded as
\begin{eqnarray}
\tan{\delta(k)} = \frac{2 \mu}{k}\,&\bigg[&
\int dr\,V(r) \sin^2{(k r)} \nonumber \\
&+& \int dr dr' V(r) \sin{(k r)} V(r') \sin{(k r')} 
G_k(r,r') \nonumber \\ &+& \mathcal{O}(V^3) \bigg] \, ,
\end{eqnarray}
i.e. the Born approximation followed by second and higher order
perturbation theory.
That is, we have written a coordinate space version of
the Lippmann-Schwinger equation.
In the expression above $G_k$ is a Green function,
which we can take to be
\begin{eqnarray}
G_k(r,r') = \frac{2\mu}{k}\,\left[ \sin{k r} \cos{k r'} \theta(r-r')
+ \cos{k r} \sin{k r'} \theta(r'-r) \right] \, .
\end{eqnarray}
If the scattering length is natural, the size of the Born and second order
term are
\begin{eqnarray}
\frac{2 \mu}{k}\,\langle V \rangle &=&
\frac{Q}{M}\,f_1(k r_c)\,
\left[
\sum_{n=0}^{\infty} c_{2n}(r_c)\,
{\left( \frac{Q}{M} \right)}^{2n}
\right] \, , \\
\frac{2 \mu}{k}\,\langle V G_0 V \rangle &=&
{\left(\frac{Q}{M} \right)}^2\,f_2(k r_c)\,
{\left[
\sum_{n=0}^{\infty} c_{2n}(r_c)\,
{\left( \frac{Q}{M} \right)}^{2n}
\right]}^2 \, ,
\end{eqnarray}
where $\langle V \rangle$, $\langle V G_0 V \rangle$ is simply a compact
notation for the first and second order of the perturbative series
and $f_1(x)$, $f_2(x)$ are functions that encode
the cut-off dependence.
If we continue we will find that for higher order perturbations we have
\begin{eqnarray}
\frac{2 \mu}{k}\,\underbrace{\langle V G_0 \dots G_0 V
\rangle}_{\mbox{$r$ iterations of $V$}}  &=&
{\left(\frac{Q}{M} \right)}^{r}\,f_r(k r_c)\,
{\left[
\sum_{n=0}^{\infty} c_{2n}(r_c)\,
{\left( \frac{Q}{M} \right)}^{2n}
\right]}^r \, ,
\end{eqnarray}
where $r$ refers to the number of insertions of the potential $V$.
The EFT expansion for the phase shift starts at $Q/M$ -- leading order (LO)
from now on -- and second order perturbation theory carries an additional
factor of $Q/M$ over the Born approximation.
Analogously each additional iteration of the potential involves
an extra power of $Q/M$.
Putting all the pieces together,
the ${\rm LO}$ calculation only contains $C_0$ at tree level,
the next-to-leading order (${\rm NLO}$) calculation requires to include
two iterations of $C_0$, the next-to-next-to-leading order
(${\rm NNLO}$) calculation contains $C_2$ at tree level and
three iterations of $C_0$.
Higher orders calculations are set up in a similar fashion.

For the large scattering length case the evaluation of
the perturbative series for the tangent of the phase shift leads to
\begin{eqnarray}
\frac{2 \mu}{k}\,\langle V \rangle &=&
f_1'(k r_c)\,
\left[ c_0(r_c) + 
\sum_{n=1}^{\infty} c_{2n}(r_c)\,
{\left( \frac{Q}{M} \right)}^{2n}
\right] \, , \\
\frac{2 \mu}{k}\,\langle V G_0 V \rangle &=&
f_2'(k r_c)\,
{\left[ c_0(r_c) +
\sum_{n=1}^{\infty} c_{2n}(r_c)\,
{\left( \frac{Q}{M} \right)}^{2n}
\right]}^2 \, , \\
\frac{2 \mu}{k}\,\underbrace{\langle V G_0 \dots G_0 V
\rangle}_{\mbox{$r$ iterations of $V$}}  &=&
f_r'(k r_c)\,
{\left[ c_0(r_c) + 
\sum_{n=1}^{\infty} c_{2n}(r_c)\,
{\left( \frac{Q}{M} \right)}^{2n}
\right]}^r \, .
\end{eqnarray}
Now the EFT expansion of the phase shift begins at order $(Q / M)^0$,
which is the ${\rm LO}$ for this power counting (i.e. the ${\rm LO}$
is defined differently for each scaling of the couplings).
It is also apparent that the ${\rm LO}$ calculation contains
all the iterations of the $C_0$ coupling.
That is, $C_0$ is non-perturbative at ${\rm LO}$.
However all the other couplings are perturbative: $C_2$ enters at tree level
at ${\rm NLO}$, $C_4$ at ${\rm N^3LO}$ and $C_{2n}$ at ${\rm N^{2n-1}LO}$.
A systematic exposition of the diagrams involved in the calculation
of the amplitudes (for natural and unnatural scattering length)
can be found in Ref.~\refcite{vanKolck:1998bw}.

\subsection{Renormalization with the One Pion Exchange Potential}

The previous analysis can be extended to the possibility that the potential
can be separated into a short and long range piece
\begin{eqnarray}
V = V_S + V_L \, .
\end{eqnarray}
The range of $V_S$ scales as $M$, while the range of $V_L$ as $Q$.
To determine the effect of a long range potential on the power counting
we follow the previous steps: introduce a cut-off and include
a contact-range potential to keep physics unchanged.
The couplings $C_{2n}$ of the contact-range potential can be calculated
from the RGE, i.e. Eq.~(\ref{eq:RGE}).
For $M r_c \gg 1$ the short-range potential vanishes and the wave functions
that enter into the RGE are solutions of  
\begin{eqnarray}
-u''_k(r) &=& k^2\, u_k(r) \quad \mbox{for $r < r_c$,} \\
-u''_k(r) + 2\mu\,V_L(r)\,u_k(r) &=& k^2\, u_k(r) \quad \mbox{for $r > r_c$,} 
\end{eqnarray}
with the boundary conditions
\begin{eqnarray}
u_k(0) &=& 0 \, , \\
u_k(r) &\to& \sin{(k r + \delta)} \quad \mbox{for $r \to \infty$,}
\end{eqnarray}
with $\delta$ the phase shift of the full potential $V_S + V_L$.
We are interested in the wave functions in the distance range
$M \gg 1/r \geq Q$.
The condition $M r \gg 1$ is necessary if we use a wave function
that is a solution of the long range potential $V_L$.
A soft cut-off --- let's say $Q r \sim 1$ --- is perfectly acceptable.
But a excessively soft cut-off of the order of $Q r \ll 1$ is not:
for this choice of the cut-off the long range potential $V_L$
vanishes and we end up with the power counting
of a pure short range potential.

Everything that is left is to calculate the wave functions
for the long range potential.
In general the form of the solution of the wave function will take the form
\begin{eqnarray}
u_k(r) = u_{a}(r; k) + c(k)\,u_{b}(r; k) \, ,
\end{eqnarray}
where $u_a$ and $u_b$ are two linearly independent solutions of $V_L$
and $c(k)$ a coefficient that selects
the particular linear combination.
If we expand the solutions in powers of $k^2$
\begin{eqnarray}
u_{a}(r;k) &=& u_{0,a}(r) + k^2\,u_{2,a}(r) + 
\sum_{n=2}^{\infty} k^{2n}\,u_{2n,a}(r) \, , \\
u_{b}(r;k) &=& u_{0,b}(r) + k^2\,u_{2,b}(r) + 
\sum_{n=2}^{\infty} k^{2n}\,u_{2n,b}(r) \, , 
\end{eqnarray}
and the coefficient $c(k)$ as
\begin{eqnarray}
c(k) &=& c_0 + c_2 k^2 + \sum_{n=2}^{\infty} c_{2n} k^{2n} \, ,
\end{eqnarray}
we end up with the following expressions for the running of
the $C_{2n}$ couplings
\begin{eqnarray}
\frac{u_{0a}' + c_0 u_{0b}'}{u_{0a} + c_0 u_{0b}}
- \frac{1}{r_c} &=& \frac{\mu}{2\pi r_c^2}\,C_0(r_c) \, , \\
\frac{c_2 u_{0b}'}{u_{0a} + c_0 u_{0b}} + 
\frac{Q_2(u_{0a}, u_{0a}', u_{0b}, u_{0b}', c_0)}{(u_{0a} + c_0 u_{0b})^2} + 
\frac{r_c}{3}
&=& \frac{\mu}{2\pi r_c^2}\,C_2(r_c) \, ,
\end{eqnarray}
plus analogous expressions for the higher order couplings, where the wave
functions and their derivatives are understood to be evaluated at $r = r_c$.
In the expression above, $Q_2$ is a polynomial of $c_0$ and
$u_{0a}$, $u_{0b}$ and its derivatives that encodes
the residual cut-off dependence.

A few general comments might be of help at this point.
First: the coefficient $c(k)$ can be thought of as the analogous of
the ERE in the presence of a long range potential.
What this means is that the set of coefficients $c_2$, $c_4$, etc.,
will scale according to inverse powers of $M$.
The exception is $c_0$, which could take any value if $V_S$ is non-perturbative.
Second: if the long range potential is perturbative, the wave functions
will coincide with the free wave functions at tree level
in perturbation theory.
The couplings will also accept a perturbative expansion, but at tree level
will coincide with the couplings of the short-range case.
Thus the power counting does not change if the long range potential
is perturbative.

In nuclear physics the longest range piece of the interaction
is the one pion exchange (OPE) potential.
This potential can be written as
\begin{eqnarray}
V_{\rm OPE}(r) = \frac{4 \pi}{M_N\,\Lambda_{\rm NN}}
\, \frac{m^3}{12 \pi} \,
\left[
W_S(r) \vec{\sigma}_1 \cdot \vec{\sigma}_2 + W_T(r) \, S_{12}(\hat{r})
\right] \, \vec{\tau}_1 \cdot \vec{\tau}_2 \, ,
\end{eqnarray}
where $\vec{\sigma}_{1(2)}$ and $\vec{\tau}_{1(2)}$ are the spin and isospin
operators acting on the nucleon 1(2), $m$ is the pion mass,
$M_N$ the nucleon mass and $\Lambda_{\rm NN}$ is a mass scale 
that characterizes the strength of the OPE potential
(its value is of the order of $300 \, {\rm MeV}$).
%
%
The tensor operator is defined as $S_{12} = 3\,\vec{\sigma}_1 \cdot \hat{r}\,
\vec{\sigma}_2 \cdot \hat{r} - \vec{\sigma}_1 \cdot \vec{\sigma}_2$,
while $W_S$ and $W_T$ refer to the spin-spin and tensor components
of the potential
\begin{eqnarray}
W_S &=& \frac{e^{-m r}}{m r} \, , \\
W_T &=& \frac{e^{-m r}}{m r}\,(1 + \frac{3}{m r} + \frac{3}{(m r)^2}) \, .
\end{eqnarray}
We will ignore the complications coming from the tensor operator
and will concentrate on the fundamentals: (i) $S_{12}$ vanishes
in the singlet and (ii) $W_S$ and $W_T$ behave as a $1/r$ and
a $1/r^3$ potential respectively.

Before analyzing the power counting with OPE it will be helpful
to comment on the role of $\Lambda_{\rm NN}$.
Notice that we have written the OPE potential as
\begin{eqnarray}
V_{\rm OPE}(r) = \frac{4 \pi}{M_N\,\Lambda_{\rm NN}} \times
P_3\,(m, \frac{1}{r})\,e^{-m r} \, ,
\end{eqnarray}
where $P_3$ is a polynomial that contains a $m^2/r$, $m/r^2$ and $1/r^3$ term,
where all the terms have three powers of $Q = \{ m \, , 1/r \}$.
The analogy with the contact-range potential is clear,
more so if we write $V_C$ as
\begin{eqnarray}
V_{\rm C}(r; r_c) = C_0(r_c) \times
\frac{1}{4\pi r^3}\,\delta(1 - \frac{r}{r_c}) \, ,
\end{eqnarray}
where we can appreciate that the ${4 \pi} / (M_N\,\Lambda_{\rm NN})$ factor
in $V_{\rm OPE}$ plays the same role as the $C_0$ coupling in $V_{\rm C}$.
That is, if we count $\Lambda_{\rm NN}$ as $M$ the OPE potential
will be perturbative~\cite{Kaplan:1998tg,Kaplan:1998we}.
On the contrary if we count $\Lambda_{\rm NN}$ as $Q$ the OPE potential
will be non-perturbative~\cite{Birse:2005um}.
We are interested in the later case: as we have already pointed out,
if OPE is perturbative the counting is the same as that of
a pure short-range potential.
The bottom line is that non-perturbative OPE goes along with
the assumption that $\Lambda_{\rm NN}$ is a light scale.
The idea that non-perturbative OPE requires the existence of an additional 
light scale (besides the obvious choices such as the external momenta,
the pion mass or the inverse of the scattering length)
was probably {\it explicitly} realized in Ref.~\refcite{Bedaque:2002mn}
for the first time.
In Ref.~\refcite{Birse:2005um} one can see how to include this scale
in the RG equations and what kind of consequence it has for the power counting.

The $1/r$ potential, which corresponds to the OPE potential
in the $^1S_0$ singlet, is the easiest to analyze.
Here we are not going to enter into the specific details of how to do
the detailed analysis.
We merely comment that the power counting is unchanged with respect to
the case where there is no long range potential~\cite{Barford:2002je}.
That is, there are two possible arrangements of the power counting:
a natural one, in which the couplings scale as $C_{2n} \sim 1/M^{2n+2}$
and an unnatural one, in which the couplings scale as $C_0 \sim 1 / (M Q)$
and $C_{2n} \sim 1/(Q^2 M^{2n})$.
The reason for that is that the $1/r$ potential is not strong enough
as to modify the behaviour of the wave functions
in the distance window $M \ll 1/r \ll \Lambda_{\rm NN}$.
Even if the strength of the potential is such as to generate a low lying
bound state, the wave functions are only substantially modified
for $\Lambda_{\rm NN} \ll 1/r \ll m$.
However this cut-off range is not of interest for power counting
because we are already making the assumption that
$\Lambda_{\rm NN} \sim m \sim Q$.

For the $^3S_1$ triplet the potential behaves as an attractive $1/r^3$ 
for $m\,r < 1$, which induces important changes
in the scaling of the couplings~\footnote{Actually this is a simplification:
the $^3S_1-{}^3D_1$ triplet is a coupled channel and the OPE is a matrix:
the tensor operator contains an attractive and repulsive eigenvalue,
and the attractive one happens to have a bigger impact
on power counting.}.
The first thing to notice is that there is not anymore a natural
and unnatural power counting.
The solutions of the wave function are all equally fine-tuned:
there is not a more natural or preferred solution
(see Ref.~\refcite{Birse:2005um} for a different conclusion though).
The reason is that the attractive $1/r^3$ potential has no unique solution
in quantum mechanics: the choice of the solution inherently depends
on the existence of short range physics, which is the only
responsible for fixing the wave function.
Every linear combination of independent wave functions is equally acceptable.
For $m\,r < 1$ the wave function can be written as~\cite{PavonValderrama:2005gu,PavonValderrama:2005wv,PavonValderrama:2005uj}
\begin{eqnarray}
u_k(r) &\propto& r^{3/4}
\sin{\left(\frac{\beta}{\sqrt{\Lambda_{\rm NN} r}} + \phi_3(k)
\right)}
\times \left[ 1 +
\mathcal{O}(\sqrt{\Lambda_{\rm NN} r}, \,k^2 r^2\, , m r) \right] \, ,
\label{eq:uk_LO}
\end{eqnarray}
where $\beta$ is a dimensionless number and $\phi_3(k)$ is a phase
-- the semiclassical phase -- that characterizes
the particular solution we are dealing with.
The value of $\phi_3(k)$ depends on the short-range physics.
The changes in the counting are the following:
\begin{eqnarray}
C_0(r_c) &\sim& \frac{1}{M Q} \, , \\ 
C_{2n}(r_c) &\sim& \frac{1}{M^{2n+3/2} Q^{1/2}} \, .
\end{eqnarray}
This result requires a careful examination of the scaling properties
of $\phi_3(k)$, which are not trivial (the derivation is not contained
here but will be included in a future publication).
As a matter of fact the counting with an attractive tensor force is more
similar to NDA than to that of a short range potential
with large scattering length.

For a repulsive singular potential the analysis is analogous
with the exception of a few details.
The wave function is
\begin{eqnarray}
u_k(r) &\propto& r^{3/4}\,\left[
\exp{\left(-\frac{\beta}{\sqrt{\Lambda_{\rm NN} r}}\right)}
+ c_3(k)\,\exp{\left(+\frac{\beta}{\sqrt{\Lambda_{\rm NN} r}}\right)} \right]
\nonumber \\
&\times& \left[ 1 +
\mathcal{O}(\sqrt{\Lambda_{\rm NN} r}, \,k^2 r^2\, , m r) \right] \, ,
\end{eqnarray}
with $c_3(k)$ a coefficient that depends on the short-range physics.
It is expected to be small and as happened with $\phi_3$
its scaling properties are important in the detailed analysis,
yet they are not trivial.
The scaling of the coupling now is
\begin{eqnarray}
C_{0}(r_c) &\sim& \frac{1}{M^{3/2} Q^{1/2}} \, , \\
C_{2n}(r_c) &\sim& \frac{1}{M^{2n+3/2} Q^{1/2}} \, .
\end{eqnarray}
That is, the only difference with the attractive case is the scaling of
the $C_0$ coupling.
However the previous scaling properties are difficult to interpret.
It is sensible to expect that the importance of short-range physics
within EFT depends on the long-range dynamics.
The attractive $1/r^3$ potential complies with this expectation:
as a consequence of the strong attraction the wave function
is enhanced at short distances, which in turn enhances
the short-range couplings.
For the repulsive $1/r^3$ potential we expect the contrary to happen,
that the size of the $C_{2n}$ couplings diminishes.
What happens is precisely the contrary, which is puzzling to say the least.

With this we have finished the discussion of power counting for the moment.
The types of power counting and the physical situations to which
they correspond are summarized in Table~\ref{tab:counting}.
Of course they are not the only types of power counting that can be built,
but they are for sure the most relevant ones for nuclear EFT.
Now I will try to show how to rederive these counting rules
with other methods. In particular I will consider
the calculation of anomalous dimensions, ultraviolet renormalizability and
residual cut-off dependence.

\begin{table}[pt]
\tbl{Power counting in the s-wave two-body system}
{\begin{tabular}{lllll} \toprule
& \multicolumn{2}{c}{$V_L = 0$} & 
\multicolumn{2}{c}{\multirow{5}{*}{$V_L \sim Q^{-1}$ ($-1/r^3$ type)}} \\
& \multicolumn{2}{c}{or} \\
& \multicolumn{2}{c}{$V_L \sim Q^{0}$ ($1/r$ or $1/r^3$ type)} \\
& \multicolumn{2}{c}{or} \\
& \multicolumn{2}{c}{$V_L \sim Q^{-1}$ ($1/r$ type)} \\
\colrule
 & $a_0 \sim 1/M$ & $a_0 \sim 1/Q$ & $V_L \sim -1/r^3$ & $V_L \sim + 1/r^3$ \\
\colrule
$C_0$ & $Q^0$ ($\rm LO$) & $Q^{-1}$ ($\rm LO$)
& $Q^{-1}$ ($\rm LO$) & $Q^{-1/2}$ ($\rm LO$) \\
$C_2\,k^2$ & $Q^2$ ($\rm N^2LO$) & $Q^0$ ($\rm NLO$)
& $Q^{3/2}$ ($\rm N^{5/2}LO$) & $Q^{3/2}$ ($\rm N^{2}LO$) \\
$C_4\,k^4$ & $Q^4$ ($\rm N^4LO$) & $Q^2$ ($\rm N^3LO$)
& $Q^{7/2}$ ($\rm N^{9/2}LO$) & $Q^{7/2}$ ($\rm N^{4}LO$)
\\
$\dots$ & $\dots$ & $\dots$ & $\dots$ \\
$C_{2n}\,k^{2n}$ & $Q^{2n}$ ($\rm N^{2n}LO$) & $Q^{2n-2}$ ($N^{2n-1}LO$)
& $Q^{2n-1/2}$ ($\rm N^{2n+1/2}LO$) & $Q^{2n-1/2}$ ($\rm N^{2n}LO$)
\\ \colrule
$\delta C_0$ & $Q^0$ (stable) & $Q^{-2}$ (unstable)
& $Q^{-1/2}$ (stable) & $Q^{-1/2}$ (stable) 
\\
\botrule
\end{tabular}}
\begin{tabnote}
Summary of the power counting for s-wave two-body scattering.
The table indicates when the coupling enters as a power of $Q$
(and the relative order in parenthesis).
In the text we have considered the case of a pionless and a pionful EFT.
For pionless the scaling of the couplings depends on the size of the scattering
length.
For pionful the scaling is identical to the pionless case if either one
of these conditions is met: (i) pion exchanges are perturbative
(ii) pion exchanges are non-perturbative but there is only
the central piece.
If the tensor piece is non-perturbative the scaling of the couplings
will be modified with respect to the previous cases.
We show this in the table by indicating whether the long-range potential
is zero ($V_L = 0$), perturbative ($V_L \sim Q^0$) or non-perturbative
($V_L \sim Q^{-1}$) and then the type of long-range potential ($1/r$
or $1/r^3$).
Finally in the last row we indicate the size of a perturbation of $C_0$,
which determines whether the power counting is
infrared stable or unstable (see discussion around Eq.~\ref{eq:dC0})
\end{tabnote}
\label{tab:counting}
\end{table}

\subsection{Power Counting and Anomalous Dimensions}

There is a very interesting simplification in the above calculations:
it is enough to take into account the cutoff dependence of $C_{2n}$
to guess its scaling~\cite{Valderrama:2014vra}.
More specifically we refer to the cut-off dependence for
$\frac{r_c}{a_0} \leq 1$ ($Q r_c \leq 1$), a condition
that will remarkably simplify the discussion below.
If we consider a two-body system with natural scattering length
the cut-off dependence of the $C_{2n}$ couplings is trivial
\begin{eqnarray}
C_{0}(r_c) &= &\frac{2\pi}{\mu}\,a_0 \, \times \,
\left[ 1 + \mathcal{O}(\frac{r_c}{a_0}) \right] \, , \\
C_{2n}(r_c) &= &\frac{2\pi}{\mu}\,a_0^2\,v_n \,\times\, 
\left[1 + \mathcal{O}(\frac{r_c}{a_0}) \right] \, ,
\quad \mbox{for $n \geq 1$,}
\end{eqnarray}
while for a system with a large scattering length we have
\begin{eqnarray}
C_0(r_c) &=&
- \frac{2\pi}{\mu}\,r_c \, \times \,
\left[ 1 + \mathcal{O}(\frac{r_c}{a_0}) \right] \, ,\\
C_{2n}(r_c) &=&
\frac{2\pi}{\mu}\,r_c^2\,v_n \, \times \,
\left[ 1 + \mathcal{O}(\frac{r_c}{a_0}) \right]
\quad \mbox{for $n \geq 1$,}
\end{eqnarray}
where to simplify the notation we have taken $v_1 = r_0 / 2$.
That is, the power-law dependence on the cut-off matches the enhancement of
the coupling.
The rule is simple: if $C_{2n}(r_c) \propto r_c^{\alpha}$ for $M \ll 1/r_c \ll Q$
the size of $C_{2n}$ for $Q r_c \to 1$ is $1 / (M^{2n-\alpha} Q^{\alpha})$,
a $M^{\alpha} / Q^{\alpha}$ enhancement.
Equivalently, in momentum space, if $C_{2n}(\Lambda) \propto \Lambda^{-\alpha}$
for $M \ll \Lambda \ll Q$
the size of $C_{2n}$ for $\Lambda \to Q$ is $1 / (M^{2n-\alpha} Q^{\alpha})$.

Why is that so?
Actually the idea can be better explained with a momentum space cut-off.
If the couplings scale as
\begin{eqnarray}
C_{2n}(r_c) \propto r_c^{\alpha} \, ,
\end{eqnarray}
with respect to the radial cut-off $r_c$,
in momentum space they will scale as~\footnote{We use the same notation
for the couplings in coordinate and momentum space: we indicate which
one we are dealing with by the argument: $r_c$ or $\Lambda$.}
\begin{eqnarray}
C_{2n}(\Lambda) \propto \Lambda^{-\alpha} \, ,
\end{eqnarray}
which simply amounts to take into account that $\Lambda \propto 1/r_c$.
This scaling property implies that the couplings follow a RGE of the type
\begin{eqnarray}
\frac{d}{d\Lambda}
\left[ \Lambda^{\alpha}\,C_{2n}(\Lambda) + \dots \right] = 0 \, ,
\end{eqnarray}
where the dots refer to corrections involving smaller powers of $\Lambda$.
For the moment we will assume that this RGE is valid
in the cut-off window $M \geq \Lambda \geq Q$.
If we ignore the dots the solution is straightforward
\begin{eqnarray}
{\Lambda_1^{\alpha}}\,{C_{2n}(\Lambda_1)} =
{\Lambda_2^{\alpha}}{C_{2n}(\Lambda_2)} \, ,
\end{eqnarray}
with $\Lambda_1$ and $\Lambda_2$ two arbitrary cut-offs.
Therefore with a boundary condition we can get the running of the couplings
for arbitrary $\Lambda$.
This boundary condition is the value of the couplings at $\Lambda = M$.
At this scale we expect the couplings to scale with $M$ (we do not expect $Q$
to play a role at high energies, which means that $M$ is the only relevant
scale), which implies
\begin{eqnarray}
C_{2n}(M) \propto \frac{1}{M^{2n+2}} \, .
\end{eqnarray}
As a consequence
\begin{eqnarray}
C_{2n}(Q) \propto \frac{1}{M^{2n+2}} \times
{\left( \frac{M}{Q} \right)}^{\alpha} \, ,
\end{eqnarray}
which is the expected enhancement for $C_{2n}$.
In short, the scaling of the coupling decides the power counting.
This idea is not new and has appeared in different contexts.
In the KSW counting~\cite{Kaplan:1998tg,Kaplan:1998we}
the hard scale can be deduced from the running of
the $C_0(\Lambda)$ coupling: the scaling of $C_0(\Lambda)$ changes
when $\Lambda$ approaches $\Lambda_{\rm NN}$,
which happens to be the hard scale in KSW~\footnote{In KSW the regularization
scale is normally referred as $\mu$ instead of $\Lambda$. It is also worth
noticing that KSW does not use a cut-off regularization, but a variant
of dimensional regularization.}.
Recently it has been applied in nuclear EFT for the analysis
of reactions on the deuteron~\cite{Valderrama:2014vra}.

There is the issue of where the RGE of the $C_{2n}$ couplings comes from,
which is related to the calculation of the power $\alpha$
that appears in it.
The starting point in Wilsonian renormalization is to include a cut-off
and then require observable quantities to be independent of the cut-off
\begin{eqnarray}
\frac{d}{d \Lambda}\,\langle \Psi | \mathcal{O} | \Psi \rangle = 0 \, ,
\end{eqnarray}
where $| \Psi \rangle$ is the wave function and $\mathcal{O}$
an operator corresponding to an observable.
Notice that here we are demanding the matrix element to be independent
of the cut-off.
Actually this condition is too strong --- observables are the square modulus of
matrix elements --- but in most situations it will work~\footnote{
An example where the phase is important can be found in the infrared
renormalization of Coulomb in proton-proton scattering of
Ref.~\refcite{Skibinski:2009ra}.}.
If we are in the cut-off window $M \geq \Lambda \geq Q$ we can substitute
the wave function and the operator by the corresponding ones in the EFT
\begin{eqnarray}
| \Psi \rangle &=& | \Psi_{\rm EFT} \rangle  \, , \\
\mathcal{O} &=& {\mathcal{O}}_{\rm EFT} \, .
\end{eqnarray}
Moreover the operator $\mathcal{O}_{\rm EFT}$ can be divided
into a contact- and finite-range piece
\begin{eqnarray}
{\mathcal{O}}_{\rm EFT} = {\mathcal{O}}_{C} + {\mathcal{O}}_{F} \, .
\end{eqnarray}
Now we can rewrite
\begin{eqnarray}
\frac{d}{d \Lambda}\,\langle \Psi_{\rm EFT} | \mathcal{O}_C | \Psi_{\rm EFT}
\rangle = -
\frac{d}{d \Lambda}\,\langle \Psi_{\rm EFT} | \mathcal{O}_F | \Psi_{\rm EFT}
\rangle
\, ,
\end{eqnarray}
which tell us that the contact-range piece has two functions:
to absorb the cut-off dependence of the finite-range piece and
to directly contribute to the matrix element.
Had we used the full wave function $| \Psi \rangle$ and
the full operator $\mathcal{O}$ instead of the EFT ones,
the contact would have only been there to absorb the cut-off dependence
(the contacts vanish for $\Lambda \to \infty$).
But within the EFT description the contacts must have a non-trivial
contribution to the observables regardless of the cut-off.
The reason is that the finite-range piece of
the EFT potential/wave function/operator does not correspond
to the fundamental potential/wave function/operator.
The bottom line is that for the contact-range operators we can distinguish
between a piece that directly contributes to observables and
a piece that absorbs cut-off dependence
\begin{eqnarray}
\mathcal{O}_C = \mathcal{O}_C^{D} + \mathcal{O}_C^{R} \, ,  
\end{eqnarray}  
where the superscript $D$ and $R$ stand for ``direct'' and ``residual''.
This distinction is analogous to the one that we made previously
for the running of the $C_{2n}$ in short-range theories.
Each of these pieces follows a different RGE
\begin{eqnarray}
\frac{d}{d \Lambda}\,\langle \Psi_{\rm EFT} | \mathcal{O}_C^D | \Psi_{\rm EFT}
\rangle &=& 0 \, , \\
\frac{d}{d \Lambda}\,\langle \Psi_{\rm EFT} | \mathcal{O}_C^R | \Psi_{\rm EFT}
\rangle &=& -
\frac{d}{d \Lambda}\,\langle \Psi_{\rm EFT} | \mathcal{O}_F | \Psi_{\rm EFT}
\rangle
\, ,
\end{eqnarray}
corresponding to their different roles within EFT.
Notice that we are assuming that the distinction between $\mathcal{O}^D$
and $\mathcal{O}^R$ exists. This is not clear (if we want the definitions
to be unambiguous), but we are only using this distinction
to simplify the arguments here.
We can write a contact-range operator $\mathcal{O}_C^D$ as a coupling
times a polynomial involving the light scales in momentum space
\begin{eqnarray}
\mathcal{O}_C^D = C(\Lambda) \times {\rm P}_{\Lambda}(Q) \, ,
\end{eqnarray}
where $C(\Lambda)$ is the coupling and ${\rm P}_{\Lambda}$ is the polynomial,
which can be regularized (hence the subscript $\Lambda$).
If we include this general form in the RGE for the ``direct'' piece
we arrive to
\begin{eqnarray}
\frac{d}{d \Lambda}\,
\left[
C(\Lambda)\,\langle \Psi_{\rm EFT} | {\rm P}_{\rm \Lambda}(Q) | \Psi_{\rm EFT}
\rangle \right] &=& 0 \, .
\end{eqnarray}
What is left is to determine the cut-off dependence of the matrix element of
the polynomial, which in general will take the form
\begin{eqnarray}
\langle \Psi_{\rm EFT} | {\rm P}_{\rm \Lambda}(Q) | \Psi_{\rm EFT} \rangle
\propto \Lambda^{\alpha} \times \left[ 1 
+ \mathcal{O}\left( \frac{Q}{\Lambda} , \frac{\Lambda}{M} \right)
\right] \, , 
\end{eqnarray}
where the form of the corrections follow from the assumption that the RGE
are valid in the region $M \geq \Lambda \geq  Q$, that is,
from the analyticity of the RGE which in turn implies
that we can write a power series on $Q / \Lambda$ and $\Lambda / M$.

The previous discussion is rather general and concerns any observable
that receives a direct, linear contribution from a contact-range operator.
In the case of two-body scattering the matrix element we are interested in
is the T-matrix
\begin{eqnarray}
\frac{d}{d\,\Lambda}\,\langle k | T_{\rm EFT} | k' \rangle = 0 \, ,
\end{eqnarray}
which is not receiving a linear contribution from the contact-range physics,
at least at first sight.
The T-matrix is the solution of the Lippmann-Schwinger equation
\begin{eqnarray}
T_{\rm EFT} = V_{\rm EFT} + V_{\rm EFT}\,G_0\,T_{\rm EFT} \, ,
\end{eqnarray}
which is a convenient way of rewriting the Schr\"odinger equation
for a scattering problem.
In EFT we can expand the T-matrix and the potential as power series
\begin{eqnarray}
T_{\rm EFT} &=&
\sum_{\nu} {\left( \frac{Q}{M} \right)}^{\nu} \hat{T}^{(\nu)} \, , \\
V_{\rm EFT} &=&
\sum_{\nu} {\left( \frac{Q}{M} \right)}^{\nu} \hat{V}^{(\nu)} \, , 
\end{eqnarray}
or more concisely as
\begin{eqnarray}
T_{\rm EFT} &=& T_{\rm LO} + \delta\,T_{\rm EFT} \, , \\
V_{\rm EFT} &=& V_{\rm LO} + \delta\,V_{\rm EFT} \, ,
\end{eqnarray}
that is, a LO contribution plus a subleading correction.
The interesting thing here is that the subleading correction to the T-matrix
is perturbative and is given by
\begin{eqnarray}
\delta\,T_{\rm EFT} &=& \langle k | (1 + T_{\rm LO} G_0)
\delta\,V_{\rm EFT} (G_0 T_{\rm LO} + 1) | k' \rangle +
\mathcal{O}\left[ {\left(\delta\,V_{\rm EFT}\right)}^2 \right] \\
&=& \langle \Psi_{\rm LO} | \delta\,V_{\rm EFT} | \Psi_{\rm LO} \rangle +
\mathcal{O}\left[ {\left(\delta\,V_{\rm EFT}\right)}^2 \right] \, ,
\end{eqnarray}
where in the second line $| \Psi_{\rm LO} \rangle$ is the LO wave function
($| \Psi_{\rm EFT} \rangle = | \Psi_{\rm LO} \rangle +
\delta\,| \Psi_{\rm LO} \rangle$).
If we ignore the iteration of $\delta\,V_{\rm EFT}$, make the separations
\begin{eqnarray}
\delta\,V_{\rm EFT} &=& \delta\,V_{C} + \delta\,V_{F} \, , \\
\delta\,V_{C} &=& \delta\,V^D_{C} + \delta\,V^R_{C} \, ,
\end{eqnarray}
and follow the steps previously described, we end up with the RGE
\begin{eqnarray}
\frac{d}{d \Lambda}\,
\langle \Psi_{\rm LO} | \delta\,V^D_{C} | \Psi_{\rm LO} \rangle = 0 \, .
\end{eqnarray}

The solution of this RGE depends on the evaluation of the matrix element
of the contact-range potential.
The details depend on the particular representation for the contacts.
For a delta-shell representation in coordinate space
\begin{eqnarray}
V_C(r; r_c) = \frac{\delta (r - r_c)}{4 \pi r_c^2} \, \sum_{n} C_{2n}(r_c) k^{2n}
\, , 
\end{eqnarray}
the evaluation is trivial 
\begin{eqnarray}
\langle \Psi_{\rm LO} | V_C | \Psi_{\rm LO} \rangle
= \frac{u_{k}^2(r_c)}{4\pi r_c^2}\,\sum_{n=0}^{\infty} C_{2n}(r_c) k^{2n}
\, ,
\end{eqnarray}
with $u_k$ the $\rm LO$ reduced wave function.
We can concentrate on the evaluation of a particular coupling $C_{2n}$,
in which case we obtain
\begin{eqnarray}
\langle \Psi_{\rm LO} | C_{2n}(r_c) k^2 | \Psi_{\rm LO} \rangle
= \frac{u_{0}^2(r_c)}{4\pi r_c^2}\,C_{2n}(r_c) k^{2n} + \mathcal{O}(k^2)
\, ,
\end{eqnarray}
where $u_0$ is the zero-energy $\rm LO$ reduced wave function.
The RGE for the coupling $C_{2n}$ reads
\begin{eqnarray}
\frac{d}{d r_c}\,\left[ \frac{u_0^2(r_c)}{r_c^2} C_{2n}(r_c) +
\dots \right] = 0 \, ,
\end{eqnarray}
which means that the running is determined by the power-law dependence
of the wave function for $Q r \leq 1$.

In a purely short-range theory
the running of the $C_{2n}$ couplings is easy to compute.
The zero-energy wave function is
\begin{eqnarray}
u_0(r) = \mathcal{A}\,\left( r -  a_0 \right) \, ,
\end{eqnarray}
where $\mathcal{A}$ is a normalization factor that is arbitrary
(it does not affect the running of $C_{2n}$).
We remind the reader that we are interested in the region $1/M \geq r \geq 1/Q$.
If the scattering length is natural ($M a_0 \sim 1$), we can take
$\mathcal{A} = 1$ and rewrite the wave function as
\begin{eqnarray}
u_0(r) =  r \times \left[ 1 + \mathcal{O}( \frac{1}{M r}) \right] \, .
\end{eqnarray}
Therefore the RGE for the couplings is
\begin{eqnarray}
\frac{d}{d r_c}\,\left[ C_{2n}(r_c) + \dots \right] = 0 \, ,
\end{eqnarray}
As a consequence $C_{2n}(Q) \sim 1/M^{2n+2}$,
in agreement with the previous determination.
If the scattering length is large ($Q a_0 \sim 1$), we set the normalization
to $\mathcal{A} = 1/a_0$ to express the wave function as 
\begin{eqnarray}
u_0(r) =  1 + \mathcal{O}( Q r )  \, ,
\end{eqnarray}
which leads to the RGE
\begin{eqnarray}
\frac{d}{d r_c}\,\left[ \frac{C_{2n}(r_c)}{r_c^2} + \dots \right] = 0 \, .
\end{eqnarray}
That is, the couplings scale as $C_{2n}(Q) \sim 1 / (M^{2n} Q^2)$.

There is a point of explain here:
the counting of $C_0$ can not always be determined with this method.
The reason is that we are calculating the scaling of the perturbative piece
of the potential.
If $C_0$ is perturbative in the first place
we will obtain the correct scaling.
This is the case in a short-range theory with natural scattering length,
where we get $C_0 \sim 1/M^2$.
On the contrary if $C_0$ is non-perturbative
the ideas presented here do not apply.
We know that $C_0$ is enhanced by $M / Q$
if the scattering length is large.
Yet the application of this method to $C_0$ is not useless: it gives us
information about the scaling of a small, perturbative change of $C_0$
\begin{eqnarray}
C_0 \to C_0 + \delta\,C_0 \, , \label{eq:dC0}
\end{eqnarray}
where $\delta\,C_0$ is enhanced as $M^2 / Q^2$.
That is, the perturbation $\delta\,C_0$ is of lower order
than the original unperturbed coupling $C_0$.
How is that so? The meaning of this enhancement for $\delta \, C_0$ is
that systems with large scattering lengths are fine-tuned~\cite{Birse:1998dk}.
A minor change in $C_0$ generates a large change in the scattering length.
In particular
\begin{eqnarray}
\delta\,C_0 = \frac{2\pi}{\mu} \frac{r_c^2}{(r_c - a_0)^2} \, \delta\,a_0 \, ,
\end{eqnarray}
which entails $\delta\,a_0 \propto a_0^2 \, \delta\,C_0$.
From the RG flow perspective the natural solution represents a
stable fixed point of the RG equations and the large scattering length
solution an unstable fixed point~\cite{Birse:1998dk}.
That is, the running of the $C_0$ coupling eventually behaves as a constant
as $r_c \to \infty$. But if the scattering length is large $C_0$ will scale
as $r_c^2$ only as far as $a_0 / r_c \geq 1$.
The scaling of $\delta\,C_0$ for the different cases that we are considering
can be consulted in Table \ref{tab:counting}.

The extension to the pionful EFT is trivial
but requires a case-by-case discussion.
If pion exchanges are perturbative (i.e. subleading),
the power counting is exactly the same as in the short-range case.
The reason is that the ${\rm LO}$ wave functions are identical
to the short-range case.
If pion exchanges are non-perturbative (i.e. leading), the power counting
depends on whether the potential is $1/r$ (central) or $1/r^3$ (tensor).
For central OPE the power counting is again as for a short-range potential
because the wave functions behave either as $1$ or as $r$ for $Q r \leq 1$.
However for attractive tensor OPE the power counting changes.
Using the wave function written in Eq.~\ref{eq:uk_LO}
we find
\begin{eqnarray}
\frac{d}{d r_c}\,\left[ \frac{C_{2n}(r_c)}{r_c^{1/2}}
\sin^2{\left[ \sqrt{\frac{\beta}{\Lambda_{\rm NN} r}} + \phi_3 \right]} +
\dots \right] = 0 \, ,
\end{eqnarray}
where the dots account for power-law corrections,
which are at least of order $\sqrt{m\,r}$ ($\sqrt{Q\,r}$).
It is worth noting that the corrections to the wave function
were computed in Ref.~\refcite{PavonValderrama:2005gu}
for the $^3S_1-{}^3D_1$ triplet.
It can be checked that they do not affect the counting.
The conclusion is that the $C_{2n}$ are bigger than expected by a factor
of $\sqrt{M/Q}$.

The title of this section makes mention of the {\it anomalous dimension}.
What is that? The concept is easy to understand.
Let us assume that we have a physical quantity
(operator, coupling, observable)
\begin{eqnarray}
A = A(Q, \Lambda, M) \, ,
\end{eqnarray}
where $A$ depends on the light scale $Q$,
on the cut-off $\Lambda$ and the hard scale $M$.
We can define several types of dimensions for $A$.
The most obvious one is the canonical dimension $d$,
which can be related to the rescaling
\begin{eqnarray}
A(\lambda Q, \lambda \Lambda ,\lambda M) = \lambda^d \, A(Q, \Lambda, M) \, ,
\end{eqnarray}
that is, the canonical dimension refers to how $A$ changes
with a change of physical units.
Another type of dimension we can define is the power counting dimension,
which refers to a rescaling of $Q$ (and $\Lambda)$ only 
\begin{eqnarray}
A(\lambda Q, \lambda \Lambda, M) = \lambda^\nu \, A(Q, \Lambda, M) \, .
\end{eqnarray}
It is important to notice that while the canonical dimension of a physical
quantity is unique, the power counting dimension is not.
Rather a physical quantity is a superposition of contributions with different
power counting dimensions
\begin{eqnarray}
A = \sum A^{(\nu)} \quad \mbox{where } \quad
A^{(\nu)}(\lambda Q, \lambda \Lambda, M) = \lambda^\nu \,
A^{(\nu)}(Q, \Lambda, M) \, .
\end{eqnarray}
The inclusion of $\Lambda$ among the things we rescale for the power counting
dimension seems counter-intuitive at first, but it is natural once we consider
the argument about the RG evolution of cut-off dependent quantities
from $\Lambda = M$ to $\Lambda = Q$.
Finally the anomalous dimension can be defined as
\begin{eqnarray}
A(Q, \lambda\,\Lambda, M) = \lambda^{a} A (Q, \Lambda, M) \, ,
\end{eqnarray}
which is exactly the kind of power-law dependence on the cut-off
that we have been studying along this section.
Thus we can restate that the anomalous dimension of a coupling is what
determines its power counting.

\subsection{Ultraviolet Renormalizability}

Wilsonian renormalization is not the most popular or widely understood
method of analyzing power counting in EFTs.
This honor corresponds to ultraviolet (UV) renormalizability,
in which contact-range couplings are included to absorb
divergences in Feynman diagrams.
Quantum electrodynamics (QED) provides a good illustration of
this idea for a quantum field theory (QFT) that only contains
marginal or relevant operators~\footnote{The importance of
a relevant (irrelevant) operator grows (decreases)
at low energies, while the size of a marginal operator
remains approximately the same regardless of energy~\cite{Polchinski:1992ed}. 
In the RGA of Refs.~\refcite{Birse:1998dk,Barford:2002je,Birse:2005um}
and also of this manuscript, the previous classification reads
as follows: for a given operator we multiply its coupling $C(\Lambda)$
times its polynomial $P_{\Lambda}(Q)$ evaluated at $Q = \Lambda$
(i.e. we take $p, p', \dots = \Lambda$ in the polynomial)
times the loop integral (which is proportional to $\Lambda$).
The running of this product for a relevant (irrelevant) operator 
behaves as a negative (positive) power of $\Lambda$ for $M > \Lambda > Q$, 
while for a marginal operator this product runs either as a constant,
as $\log{\Lambda}$ or more generally as something that is not power-law.
},
i.e. what is traditionally known as a {\it renormalizable} QFT.
At this point it is important to mention
that nowadays --- after the discovery of EFTs ---
renormalizability is understood in a broader sense.
Yet for EFTs the application of this principle is simple:
we begin by considering the matrix element of an EFT operator
between EFT wave functions
\begin{eqnarray}
\langle \Psi_{\rm EFT} | \mathcal{O}_{\rm EFT} | \Psi_{\rm EFT} \rangle \, .
\end{eqnarray}
The operator contains a finite- and a contact-range piece. For the moment
we will ignore the contact-range piece because we want to use
these operators to remove divergences in the finite-range piece.
Thus we consider
\begin{eqnarray}
\langle \Psi_{\rm EFT} | \mathcal{O}_{\rm F} | \Psi_{\rm EFT} \rangle \, .
\end{eqnarray}
Now we expand this matrix element in powers of $Q / M$ as before. To simplify
the analysis we only take into account the ${\rm LO}$ wave functions
\begin{eqnarray}
\sum_{\nu}\,\langle \Psi_{\rm LO} | \mathcal{O}_{\rm F}^{(\nu)}
| \Psi_{\rm LO} \rangle + \mathcal{O}\left( \delta \, \Psi^{(\nu)} \right) \, ,
\end{eqnarray}
where we expect the contributions coming from the subleading corrections
to the wave function to be inessential for the analysis.
In the final step we isolate the $\nu$-th order contribution,
include a cut-off $\Lambda$ and check whether the matrix element
\begin{eqnarray}
{\langle \Psi_{\rm LO} | \mathcal{O}_{\rm F}^{(\nu)} | \Psi_{\rm LO} \rangle
}_{\Lambda}
\end{eqnarray}
is finite for $\Lambda \to \infty$.
If not, we include contact-range contributions until the matrix element
\begin{eqnarray}
\lim_{\Lambda \to \infty}
{\langle \Psi_{\rm LO} | \mathcal{O}_{\rm F}^{(\nu)} + \mathcal{O}_{\rm C}^{(\nu)}
| \Psi_{\rm LO} \rangle}_{\Lambda}
\end{eqnarray}
is finite.
If a divergences requires the inclusion of a new contact at order $\nu$,
the contact is counted as being of this order.

We can illustrate the idea in non-relativistic scattering,
where the relevant matrix element is
\begin{eqnarray}
\langle \Psi_{\rm LO} | V_{\rm F}^{(\nu)} | \Psi_{\rm LO} \rangle =
\int_{r_c}\,dr\, V_{\rm F}^{(\nu)}(r)\,u_{k}(r)^2 \, ,
\end{eqnarray}
where $u_k(r)$ represents the ${\rm LO}$ reduced wave function
and $r_c$ ($\propto 1/\Lambda$) is the radial cut-off.
In the formula above a contribution to the finite-range potential is said
to be of order $\nu$ when it contains $\nu$ powers of the light scales
in the momentum space representation
\begin{eqnarray}
\langle p' | V_{\rm F}^{(\nu)} | p \rangle \propto
\frac{Q^{\nu}}{M^{\nu + 2}}\,f(\frac{Q}{Q'}) \, ,
\end{eqnarray}
where $Q$ includes $p$, $p'$, the pion mass $m$ in pionful nuclear EFT
and/or other scales depending on the particular EFT we are dealing with.
The expression $Q / Q'$ refers to an arbitrary ratio of light scales
(for instance, $p / m$ and $p' / m$ in nuclear EFT) and $f$ is a 
non-polynomial function that we must compute from the EFT Lagrangian,
but which exact form is not important at this point.
If we Fourier-transform this expression into coordinate space (and
assume for simplicity that the potential is local), we find
\begin{eqnarray}
V_{\rm F}^{(\nu)}(r) \propto
\frac{1}{M^{\nu + 2}\,r^{\nu + 3}}\,f'(\frac{Q}{Q'}) \, ,
\end{eqnarray}
where $Q / Q'$ refers to $m\,r$ in nuclear physics.
The point is that we know the UV behaviour of the EFT potential.
Provided we have the ${\rm LO}$ wave functions
we can analyze the matrix element
\begin{eqnarray}
\langle \Psi_{\rm LO} | V_{\rm F}^{(\nu)} | \Psi_{\rm LO} \rangle \propto
\int_{r_c}\,dr\, \frac{u_{k}(r)^2}{r^{\nu +3}} \, ,
\end{eqnarray}
for divergences and decide which contacts to include.

The complete analysis can be found
in Refs.~\refcite{Valderrama:2009ei,Valderrama:2011mv}.
Here we merely comment on the results.
If the ${\rm LO}$ wave function comes from the $1/r$ potential,
perturbative and Wilsonian renormalization lead to identical power countings.
This is also true if the ${\rm LO}$ wave function comes
from a purely contact-range potential.
On the contrary if the ${\rm LO}$ potential is of the $1/r^3$ type
and attractive there is a small, yet significant difference
between perturbative and Wilsonian renormalization.
Removing the divergences only requires the $C_{2n}$ to enter at order
$\nu = (5 n-1)/2$, in contrast with $\nu = 2 n - 1/2$ from RGE.
The apparent scaling of the couplings is thus
$C_{2n} \sim Q^{(n-1)/2} / M^{(5 n+3) / 2}$,
i.e. an extra suppression of ${(Q/M)}^{n/2}$
with respect to the Wilsonian renormalization
value $C_{2n} \sim 1 / (M^{2 n + 3/2} Q^{1/2})$.
If the ${\rm LO}$ $1/r^3$ potential is repulsive the matrix elements
for scattering are always finite and no contact interaction
is required. However we will not discuss this problem here.
Back to the attractive $1/r^3$ potential the reason for the mismatch
probably has to do with the $k^2$ expansion of the ${\rm LO}$ wave
function, which induces a contamination of ${( \Lambda_{\rm NN} )}^{n/2}$
into the $C_{2n}$ coupling.
In fact the $k^2$ expansion of the ${\rm LO}$ wave function reads
\begin{eqnarray}
u_k(r) &=& r^{3/4} \sin{\left(\frac{\beta}{\sqrt{\Lambda_{\rm NN} r}}
+ \phi_3
\right)}
\nonumber \\ &\times&
\left[ c_0 + c_2 \, (k r)^2 \sqrt{\Lambda_{\rm NN} r} +
c_4 \, (k r)^4 (\sqrt{\Lambda_{\rm NN} r})^2 + \dots
\right]
\end{eqnarray}
with $\beta$, $c_0$, $c_2$, $c_4$, etc. numerical coefficients and
where $\phi_3$ now is independent of energy~\footnote{When we wrote
the $1/r^3$ wave functions for the RG equations we included an energy-dependent
semiclassical phase (see Eq.~\ref{eq:uk_LO}),
instead of an energy-independent one like here.
The reason is that here we are writing the ${\rm LO}$ wave functions,
in which only the energy-independent $C_0$ operator contributes,
while there we were writing the generic solution of
the $1/r^3$ potential with arbitrary short-range physics.}.
We can see that each two powers of $k$ imply half a power of
$\Lambda_{\rm NN}$ 
As a consequence the couplings that make the matrix element of the potential
finite are not the standard $C_{2n}$'s but implicitly contain $n$ half integer
powers of $\Lambda_{\rm NN}$.
That is, they have a different operator structure:
$\Lambda_{\rm NN}^{n/2} k^{2n}$ instead of $k^{2n}$.
In the same way that it is useful to make the distinction
\begin{eqnarray}
C_0 \quad \mbox{versus} \quad D_2\,m_{\pi}^2 \, ,
\end{eqnarray}
we can also write
\begin{eqnarray}
C_{2n}\,k^{2n} \quad \mbox{versus}
\quad E_{2n}\,k^{2n}\,\Lambda_{\rm NN}^{n/2}\, ,
\end{eqnarray}
to make the different structure of these couplings explicit.
In this notation the $E_{2n}$'s happen to be enhanced by $Q^{1/2}$,
just as the $C_{2n}$'s.
However the drawback of this explanation is that
unlike the $D_2$ coupling, the proposed $E_{2n}$ couplings
do not have a clear interpretation at the lagrangian level.

\subsection{Residual Cut-off Dependence}

The analysis of the residual cut-off dependence of the matrix elements is
another method for determining
the power counting~\cite{Long:2011qx,Long:2011xw,Long:2012ve}.
First we will review the theoretical basis for this idea:
for that we consider a matrix element for which all UV divergences
have been removed at the arbitrary order $\mu$.
The matrix element still contains a residual cut-off dependence that
vanishes for $r_c \to 0$:
\begin{eqnarray}
\langle \Psi_{\rm LO} | V_{\rm F}^{(\mu)} + V_{\rm C}^{(\mu)} |
\Psi_{\rm LO} \rangle = V_0 + V_a \, r_c^a + \dots \, ,
\end{eqnarray}
where $V_{\rm C}^{(\mu)}$ refers to the contact-range potential
that renormalizes the order $\mu$ calculation,
while $V_0$ and $V_a$ are coefficients.
The point is that the residual cut-off dependence indicates that
the next new higher-order coupling of the contact-range potential
enters at order $\mu + a$.
What is the reason for that? Let us assume that the next divergence indeed
enters at order $\mu + a$. The softest divergence that we are expected
to find in the matrix elements of the potential is logarithmic, thus
\begin{eqnarray}
\langle \Psi_{\rm LO} | V_{\rm F}^{(\mu + a)} + {V'}_{\rm C}^{(\mu)} |
\Psi_{\rm LO} \rangle \propto \log{r_c} \, ,
\end{eqnarray}
where the previous matrix element is the one for the finite-range potential
at order $\mu + a$ plus the number of contact-range couplings
that is expected at order $\mu$.
Notice that the value of these couplings change order-by-order,
but their number only changes at the order at which a new $C_{2n}$
coupling is included: we have written ${V'}_{\rm C}^{(\mu)}$
with a prime to indicate this fact.
As the divergence of the finite-range potential is
$V_F^{(\nu)}(r) \sim 1/r^{3+\nu}$,
it is not difficult to infer that going one order down translates into
a residual cut-off dependence of $r_c$ while going one order up
gives rise to a $1/r_c$ divergence. 
Equivalently, if we move $a$ orders down the expansion 
the residual cut-off dependence will be $r_c^a$,
from which the previous conclusion about power counting follows.

After this an example might be the best way to illustrate the method.
The easiest one is that of a contact-range theory
with a large scattering length.
If we solve $k\,\cot{\delta}$ at ${\rm LO}$ for the delta-shell short-range
potential that we have been using, we obtain the result
\begin{eqnarray}
k\,\cot{\delta} = - \frac{1}{a_0} + 
\left[ \frac{2}{3}\,r_c - \frac{1}{3}\,\frac{r_c^2}{a_0} \right]\, k^2 + 
\mathcal{O}(k^4) \, ,
\end{eqnarray}
where the residual cut-off dependence if of order $r_c$. 
The conclusion is that the next counterterm is one order below $C_0$.
That is, $C_2$ enter at $\rm NLO$.
If we now proceed to compute $k\,\cot{\delta}$ at ${\rm NLO}$
we find~\footnote{The ${\rm NLO}$ calculation includes $C_2$
at first order perturbation theory. Otherwise the residual
cut-off dependence will be different.}
\begin{eqnarray}
k\,\cot{\delta} = - \frac{1}{a_0} + \frac{1}{2}\, r_0 \,k^2 + 
\left[ \frac{1}{6}\,r_0\,r_c^2 + \mathcal{O}(r_c^3) \right]\, k^4 +
\mathcal{O}(k^6) \, .
\end{eqnarray}
The residual cut-off dependence is now of order $r_c^2$:
$C_4$ enters two orders below $C_2$, that is ${\rm N^3LO}$.
Strictly speaking residual cut-off dependence is a constructive process
and we can use it to determine the location of only the next coupling
that enters in the theory, but not more.
If we want to find the order of $C_6$ we must first compute
the ${\rm N^3LO}$ amplitudes that contain $C_4$ and
from this extract the residual cut-off dependence.
Alternatively, we can always rely on the natural expectation
that $C_6$ should enter two orders below $C_4$.

Finally it is interesting to check the predictions of this idea
for the tensor force.
On general grounds we expect the cut-off dependence of a $\rm LO$ calculation
of the phase shift with attractive tensor OPE
to be~\cite{PavonValderrama:2007nu}
\begin{eqnarray}
\frac{d\,}{d\,r_c}\,\delta_{\rm LO} \propto r_c^{3/2} \, ,
\end{eqnarray}
which after integration leads to a residual dependence of $r_c^{5/2}$.
This indicates that $C_2$ enters at $\rm N^{5/2}LO$
in agreement with the previous determinations.

\subsection{Power Counting and Wilsonian Renormalization}

The central point of this section has been to review how we can derive
power counting in Wilsonian renormalization.
The application of renormalization group analysis (RGA) to nuclear EFT,
though sometimes considered a bit arcane,
can lead to interesting insights.
To illustrate the idea we have taken non-relativistic s-wave scattering
as an example and shown in detail how to derive well-known facts
about power counting in the two-body sector
that we review in Table \ref{tab:counting}.
We have used two equivalent RG formulations.
The first is the standard one in which the starting point is a 
``fundamental theory'': we include a cut-off in the theory and
then evolve it from the ultraviolet to the infrared.
As a result we find a physical theory --- the EFT --- that is equivalent
to the fundamental theory at low energies.
The EFT incorporates the familiar counting rules that we already know,
for instance the enhancement of the couplings
when the scattering length is large.
The second is a more streamlined formulation in which we do not directly
evolve the EFT from the fundamental theory and instead use a convenient
shortcut to determine the size of the effective couplings.
This shortcut is the calculation of the anomalous dimension of the couplings,
which turns out to be relatively easy, at least in the two-body case.
As we will see, this is also the case for reactions of external probes
on two-body states and for the three-body problem in pionless.

Other important point is the relationship between RGA and
more standard techniques of determining the power counting.
By more standard techniques we refer to ultraviolet renormalization
and residual cut-off dependence.
In principle we expect all derivations to be equivalent.
In practice this equivalence has to be shown by means of concrete calculations.
The results indicate the direct equivalence with RGE in the absence of
singular pion exchanges, i.e. in the absence of the tensor force.
If the tensor force is present there is an apparent contradiction though:
the $C_{2n}$ couplings seem to be more demoted in ultraviolet renormalization
than in RGA or in residual cut-off dependence.
This disagreement can be explained as a contamination of the $C_{2n}$ couplings
with the $\Lambda_{\rm NN}$ scale. In other words, what we call $C_{2n}$ in
ultraviolet renormalization is not really the $C_{2n}$ coupling,
but rather a coupling with the structure $E_{2n}\, k^{2n} \Lambda_{\rm NN}^{n/2}$
instead of the expected $C_{2n}\,k^{2n}$.
This distinction is in fact analogous to the one
that is usually made between $C_0$ and $D_2\,m_{\pi}^2$.

There are a few open problems that we have not addressed though.
The most obvious example is the power counting of the triplet channels
where the OPE potential is a repulsive $1/r^3$.
The RG evolution of the couplings indicates that, with the exception of $C_0$, 
the scaling of the couplings is identical to that of
the attractive $1/r^3$ potential.
This conclusion agrees with a previous RGA of
the OPE potential~\cite{Birse:2005um},
but it is counterintuitive to say the least.
If the long-range physics is repulsive we expect that the short-range
physics will play a lesser role at low energies because the repulsive
long-range physics acts as a potential barrier.
That is why some authors prefer to use naive dimensional
analysis in this case~\cite{Valderrama:2011mv,Long:2011xw}.
Other problem that is related to the previous one is what happens
with coupled channels such as the $^3S_1-{}^3D_1$ deuteron channel.
In this latter case three different countings have been
proposed~\cite{Birse:2005um,Valderrama:2009ei,Long:2011xw}.

Even in the attractive triplet channels there are two proposals about
the scaling of the $C_0$ coupling: does it enter at $\rm LO$ ($Q^{-1}$)
or at $\rm N^{1/2}LO$ ($Q^{-1/2}$)?
The RGA of Birse~\cite{Birse:2005um} assumes that $C_0$ is
$\rm N^{1/2}LO$ in the attractive triplet channels.
But here we have taken the view that $C_0$ is $\rm LO$:
it has to be there because the $\rm LO$ wave function of a non-perturbative
attractive triplet is not well defined
without the inclusion of short-range physics.
We find it worth noticing that the perturbation $\delta\,C_0$ of
this coupling is $\rm N^{1/2}LO$($Q^{-1/2}$),
which is where $C_0$ is predicted to be by Birse's RGA~\cite{Birse:2005um}.
That is, the cause of the disagreement seems to be
that Ref.~\refcite{Birse:2005um} overlooks
the presence of short-range physics in the $\rm LO$ wave functions:
the $C_0$ coupling is implicit in the choice of a semiclassical phase,
i.e. the choice of $\phi_3$ in Eq.~(\ref{eq:uk_LO}).

One last problem is the scaling of the $C_{2n}$ couplings in the $^1S_0$
singlet. The standard counting~\cite{Kaplan:1998tg,Chen:1999tn}
is that the piece of $C_{2n}$ that carries physical information
enters at $\rm N^{2n-1}LO$: $C_2$ at $\rm NLO$,
$C_4$ at $\rm N^3LO$, $C_6$ at $\rm N^5LO$, etc.
Long and Yang~\cite{Long:2012ve} get to a different conclusion instead:
the $C_{2n}$'s enter at $\rm N^{n}LO$:
$C_2$ at $\rm NLO$, $C_4$ at $\rm N^2LO$, $C_6$ at $\rm N^3LO$, etc.
The conclusion is a bit puzzling: according to Ref.~\refcite{Long:2012ve},
dimensional regularization with minimal subtraction leads to a stronger
enhancement of the $C_{2n}$ couplings than cut-off regularization

Other aspect to discuss is the interpretation of the cut-off in EFT.
The RG equations use a cut-off in the region $M \geq \Lambda \geq Q$.
This raises the question of whether the cut-off should stay below
the breakdown scale, as happens in the RGA.
The answer is {\it not necessarily}.
The RG equations are formulated with the limits $M \geq \Lambda $
and $\Lambda \geq Q$ in mind to uncover
the scaling of the couplings
This is a formal requirement to make the analysis easier,
not a practical requirement in EFT calculations.
EFTs are RG-invariant: the cut-off does not appear in the observable quantities
that we compute, only in the intermediate calculations leading
to the EFT predictions.
That is, the cut-off is kept low in RGA with the intention of making
the scaling of the contact-range couplings as obvious as possible.
Once the RGA is done
the only constraints about the size of the cut-off are practical ones.

One of these constraints is the existence of residual cut-off independence.
In most EFT calculations we do not include all the couplings
that are required to achieve exact RG independence.
We only include the couplings that carry physical information
at the order we are considering.
There are two reasons for doing this: first, exact RG independence is not
well-defined if we are making calculations in an EFT at a given order.
The systematic EFT error is always present and RG independence
must be understood within this error.
The second reason is that exact RG independence requires the inclusion
of what we have called here the redundant couplings, i.e. the $C_{2n}^R$
piece of the couplings in the RG equations.
These redundant operators can be calculated and included explicitly
in a few specific cases: the KSW counting~\cite{Kaplan:1998tg,Kaplan:1998we}
and pionless EFT with PDS~\cite{Chen:1999tn}.
But on more general cases this is unpractical and not really necessary.
It is easier to keep the residual cut-off dependence
under control by a judicious choice of the cut-off.
For this condition to be true it is usually enough for $\Lambda$
to be of the order of the hard scale, though the exact details
will depend on the regulator.

Other important thing is to stress that a power counting is merely
an ideal organization of the size of the interactions of a theory.
They are derived under the assumption that the scale separation is large
and that we can clearly classify all scales either as soft ($Q$) or hard ($M$).
However the real physical world is not necessarily like that. What do we do
if we have a two-body system with a scattering length that
is neither small nor large?
The point is that what we obtain with RGA is just an approximation
to a more complex reality.
In particular other power countings are possible beyond
the ones we have discussed here. 
For instance in Ref.~\refcite{vanKolck:1998bw} van Kolck
developed a counting for two-body systems
in which the scattering length is tiny.
Other possibility is when both the scattering length and
the effective range are large, a case which can be useful
for the description of low lying s-wave resonances or
even for the $^1S_0$ singlet to improve
the convergence~\cite{Long:2013cya}.
That means that we are entitled to curb the counting rules in view of practical
physical information of the system. 
The limit is theoretical consistency: the EFT must be equivalent to
the fundamental theory at low energies, which means
that renormalizability must be respected.

\section{Beyond the Two-Body Problem}

The principles of renormalization work in the same way
for operators different than the two-body potential.
The advantage of calculating the anomalous dimension is that
we can extend the idea seamlessly to any other problem.
The point is to have a coupling and a polynomial contact-range operator,
to evaluate their matrix element and
to demand RG invariance at the end,
\begin{eqnarray}
\frac{d}{d\Lambda}
\langle \Psi_{\rm LO}' | O_C | \Psi_{\rm LO} \rangle = 0 \, .
\end{eqnarray}
Actually the whole process amounts to nothing more than following a recipe.
The only thing we have to do is to choose the wave functions and the operators
that are appropriate for the particular physical process we are studying.

\subsection{External Probes and Power Counting}

Let us consider the case of a reaction involving a external probe
and the deuteron (or more generally the two-nucleon system).
In this case the contact-range operators of the theory involve two nucleons
and one (or more) external fields.
The external fields we are interested in are pions, photons and neutrinos.
In principle the initial and final wave functions are the product of a
two-nucleon wave function and the wave function of zero, one or
more external probes
\begin{eqnarray}
| \Psi_{\rm EFT} \rangle = | \Psi_{\rm NN} \, , \, 
\{ \phi_i(\vec{q_i}) \} \rangle \, ,
\end{eqnarray}
where $\{ \phi_i \}$ refers to the probes, with the index $i = 0, 1 \dots n$.
The contact-range operator involves a coupling and a polynomial
of the momenta of the nucleons and the external probes.
In the plane wave basis it reads
\begin{eqnarray}
\langle \vec{p}\,' \, , \, \{ \phi_j'(\vec{q}_j\,') \} |
O_C | \vec{p} \, , \, \{ \phi_i(\vec{q}_i) \} \rangle
= C(\Lambda) \times P_{\Lambda}(\vec{p},\vec{p}\,',\vec{q}_i,\vec{q}_j\,') \, ,
\end{eqnarray}
where $\vec{p}$ ($\vec{p}'$) is the center-of-mass momentum of the initial
(final) two-nucleon system and $\vec{q}_i$($\vec{q}_j\,'$) the momenta of
all the incoming (outgoing) probes involved in the operator.
In general the polynomial $P_{\Lambda}$ will involve spin and isospin degrees
of freedom, but for the moment we will ignore them.

How does one evaluate the matrix elements?
If we consider that the wave functions of the external probes are plane waves,
the evaluation of the contact-range operator yields
\begin{eqnarray}
\langle \Psi_{\rm EFT}' | O_C | \Psi_{\rm EFT} \rangle = 
C(\Lambda)\,
\langle \Psi_{\rm NN}' | 
P_{\Lambda}(\vec{p},\vec{p}\,',\vec{q}_i,\vec{q}_j\,')
| \Psi_{\rm NN} \rangle \, ,
\end{eqnarray}
which at the end involves the matrix element of a polynomial between
the initial and final two-nucleon wave functions.
That is, we end up with the same type of matrix elements
as in two-nucleon scattering.
The calculation of the RGE is done as in the case of two-nucleon scattering,
except for two differences.
The first difference is that the polynomial contains new pieces that were
not present before: the momenta of the external probes.
However they factor out of the matrix element and do not contribute
to the RGE evolution of the couplings.
The second difference is that the initial and final two-nucleon wave
functions can be different.
While in two-nucleon scattering the initial and final states involve
the same scattering channel ($^1S_0$, $^3S_1-{}^3D_1$, $^3P_0$...),
this is not true in general for a reaction.
The reason is that the external probe carries quantum numbers, which means
that the scattering channel can change in the reaction.
We can have transitions from $^3S_1-{}^3D_1$ to $^1S_0$ and other combinations.
As a consequence the RGE ends up being
\begin{eqnarray}
\frac{d}{d r_c} \left[ \frac{u(r_c) u'(r_c)}{r_c^2} C(r_c) + \dots
\right] = 0 \, ,
\end{eqnarray}
where $u$ and $u'$ are the reduced wave functions of
the initial and final states.

The evaluation of the anomalous dimension for a coupling $C$ depends
on the channels involved in the reaction.
If we consider a short-range theory (or a theory containing a long-range
potential that is perturbative), the $^1S_0$ and $^3S_1$ partial waves
behave in exactly the same way.
The outcome is an enhancement of $M^2/Q^2$ in any contact-range operator
involving these partial waves.
If we consider a pionful EFT with non-perturbative pions the enhancement
depends on the partial waves involved: the presence of the $^1S_0$ partial
wave will elicit a $M/Q$ enhancement, while the $^3S_1$ a $(M/Q)^{1/4}$.
These factors must be multiplied: a $^3S_1 \to {}^3S_1$ transition involves
a $(M/Q)^{1/4} \times (M/Q)^{1/4} = (M/Q)^{1/2}$  enhancement,
a $^3S_1 \to {}^1S_0$ a $(M/Q)^{1} \times (M/Q)^{1/4} = (M/Q)^{7/4}$
and a $^1S_0 \to {}^1S_0$ a $(M/Q)^{1} \times (M/Q)^{1} = (M/Q)^{2}$.
In short, the result is surprisingly simple.

The extension to P-waves, though not derived in the present review,
is worth a brief comment: while in pionless they do not entail any enhancement,
in pionful a transition involving a $^3P_0$ partial wave will generate
a $(M/Q)^{5/4}$ enhancement~\footnote{For P-waves the enhancements are
bigger than in S-waves. The reason is that the P-wave contacts are
initially more suppressed, which also means that there is more
room for increasing the size of the couplings.}.
In contrast the $^1P_1$ does not involve any enhancement,
the $^3P_1$ most probably not (though it is an instance of a 
repulsive $1/r^3$ and thus open to discussion) and
the $^3P_2$ probably generates the same enhancement as the $^3P_0$.
Putting the pieces together we see that the two transitions
that will be most enhanced in pionful are $^1S_0 \to {}^3P_0$
and $^1S_0 \to {}^3P_2$
by a factor $(M/Q)^{1} \times (M/Q)^{5/4} = (M/Q)^{9/4}$,
followed by $^3S_1 \to {}^3P_0$ and $^3S_1 \to {}^3P_2$
by a factor $(M/Q)^{1/4} \times (M/Q)^{5/4} = (M/Q)^{3/2}$.
In electromagnetic processes $^1S_0 \to {}^3P_0$ is forbidden, but
the others can appear as magnetic quadrupole and electric dipole transitions.
Curiously these S- to P-wave transitions can be very interesting
if one considers parity violation~\cite{Zhu:2004vw,deVries:2015pza}
or parity plus time-reversal violation,
in which case they might contribute to the electric dipole moments of
light nuclei~\cite{deVries:2011an} (either with external probes or
as a part of the potential).

\subsection{Electroweak Reactions on the Deuteron}

The application of the previous ideas to electroweak reactions is mostly direct
except for the symmetry constraints of each particular case.
Electromagnetic processes respect gauge symmetry and as a consequence
also charge conservation.
This will have an impact on which are the allowed contact two-body currents
in a reaction.
The lagrangian interaction term of a reaction involving a single photon
takes the general form $A_{\mu} J^{\mu}$, with $A_{\mu}$ the photon field
and $J^{\mu}$ the electromagnetic current and
$\mu = 0,1,2,3$ a Lorentz index.
For a matrix element involving initial and final two-nucleon states
that are on the mass shell, $J^{\mu}$ obeys the Ward identity
\begin{eqnarray}
q_{\mu} \, \langle \Psi' | J^{\mu}(q) | \Psi \rangle = 0 \, . 
\end{eqnarray}
We can distinguish between two pieces of the current: the longitudinal piece,
which is parallel to the moment of the photon $\vec{q}$,
and the transversal piece, which is perpendicular to $\vec{q}$.
Reactions where the external probe is a photon --- deuteron photodisintegration
and radiative capture of neutron by protons ($\gamma d \to n p$  and
$n p \to d \gamma$) --- depend on the transversal part of
the current and are not constrained by gauge symmetry.
However there is the indirect constraint that the two-body current operator
is transversal to the photon momentum, which entails that the lowest
dimensional operator we can build is
\begin{eqnarray}
\langle \vec{p}\,' | {\vec{J}\,}^T_{2B}(\vec{q}) | \vec{p}\, \rangle =
M(\Lambda)\,\vec{\beta} \times \vec{q} \, ,
\end{eqnarray}
where $\beta$ a pseudovector that encodes the spin and isospin dependence.
The operator contains one power of the external momentum $\vec{q}$,
that is, one power of $Q$.
For determining at which order in the EFT expansion this operator enters
we have to compare with the one-body current operator, which
in the Breit frame~\footnote{The Breit frame is equivalent to taking the zero
component of the photon quadrimomentum equal to zero,
i.e. there is no energy transfer.
In more practical terms this means that if we have an incoming photon with
3-momentum $\vec{q}$ then the total 3-momentum of the incoming and outgoing
two-nucleon system is $\vec{P} = -\vec{q}/2$ and $\vec{P}\,' = +\vec{q}/2$.
}
reads
\begin{eqnarray}
\langle \vec{p}\,' | \vec{J}_{1B}(\vec{q}) | \vec{p}\, \rangle &=&
\left[ e\,\frac{\vec{p}\,' + \vec{p}}{2 M_N} + i \hat{\mu}_B \times \vec{q}
\right]\,
\delta (\vec{p}\,' - \vec{p} - \frac{\vec{q}}{2}) \, , 
\end{eqnarray}
that scales as $Q^{-2}$ (because the Dirac delta counts as $Q^{-3}$).
In the expression above $e$ is the charge of the two-nucleon state,
$\vec{p}$ and $\vec{p}\,'$ the two-nucleon center-of-mass initial
and final momenta, $M_N$ the nucleon mass and
$\hat{\mu}_B$ the magnetic moment operator.
In NDA the coupling $M(\Lambda)$ scales as $1/M^4$ and
the two-body contact-range current
enters at ${\rm N^3LO}$ compared to the one-body current.
However in nuclear physics this two-body current is sandwiched between
a $^1S_0$ and $^3S_1$ partial wave and that changes the anomalous
dimension of the coupling.
As a consequence of the enhancements that we analyzed previously,
in pionless~\cite{Chen:1999tn} the contact current enters at ${\rm NLO}$,
while in pionful~\cite{Valderrama:2014vra} enters at ${\rm N^{7/4}LO}$.

Now we consider the deuteron form factors.
They are important for elastic electron-deuteron scattering,
which is a reaction mediated by a virtual photon.
That is, the longitudinal as well as the transversal current
will have to be taken into account.
The deuteron is not a point particle: its response to a virtual photon
is described with form factors.
As the angular momentum of the deuteron is $J=1$
there are three independent form factors~\footnote{Assuming parity
and time-reversal invariance. Otherwise there will be more form factors.}:
the charge, the magnetic and the quadrupole form factor.
In the Breit frame
they are defined as~\cite{Phillips:1999am,Gilman:2001yh,Valderrama:2007ja}
\begin{eqnarray}
G_C(\vec{q}) &=& \frac{1}{3 e}\,\sum_{m_d = -1}^{+1}\,
\langle \Psi_d (1 m_d) | J_0(\vec{q}) | \Psi_d(1 m_d) \rangle \, , 
\\
G_M(\vec{q}) &=& - \frac{1}{e \sqrt{2 \eta}}\,
\langle \Psi_d (1 1) | J_1(\vec{q}) + i J_2(\vec{q}) | \Psi_d(1 0) \rangle \, ,
\\
G_Q(\vec{q}) &=& \frac{1}{2 e \eta M_d^2}\,\left[
\langle \Psi_d (1 0) | J_0(\vec{q}) | \Psi_d(1 0) \rangle -
\langle \Psi_d (1 1) | J_0(\vec{q}) | \Psi_d(1 1) \rangle
\right] \, ,
\end{eqnarray}
where $| \Psi_d(1 m_d) \rangle$ refers to the deuteron wave function with
the third component of the total spin being $m_d$,
$\eta = Q^2 / (4 M_d^2)$ where $Q^2 = |\vec{q}|^2 - |q_0|^2$,
with $q$ the 4-momentum of the virtual photon
and $M_d$ is the deuteron mass.
The 3-momentum of the photon is taken to be in the $i = 3$ direction,
i.e. $\vec{q} = (0,0,q)$.
That is why the matrix elements of the $J_3$ component of the current
are not considered above: these matrix elements are related to
the matrix elements of $J_0$ by means of the Ward identity.

The charge and quadrupole form factors depend on the charge current $J_0$.
The lowest dimensional contact operator contributing to $J_0$ is in principle
\begin{eqnarray}
\langle \vec{p}\,' | J^0_{2B}(\vec{q}) | \vec{p}\, \rangle = C(\Lambda) \, ,
\end{eqnarray}
but this operator is forbidden by charge conservation,
the most direct consequence of gauge symmetry.
The reason is that it gives a non-zero contribution to the deuteron charge.
The lowest dimensional operators compatible with charge conservation are
\begin{eqnarray}
\langle \vec{p}\,' | J^0_{2B}(\vec{q}) | \vec{p}\, \rangle =
D(\Lambda)\,\vec{q\,}^2 +
Q(\Lambda)\,\left[ 3 (\vec{S} \cdot \vec{q})^2 - \vec{q\,}^2 \right]\, ,
\end{eqnarray}
with $\vec{S}$ the spin operator of the deuteron.
The couplings $D(\Lambda)$ and $Q(\Lambda)$ represent a direct contribution to
the deuteron charge radius and quadrupole moment respectively.
Their size is $1/M^4$ in NDA.
The deuteron wave functions induce an anomalous dimension for $D(\Lambda)$
and $Q(\Lambda)$, which scale as $1/M^2 Q^2$ in pionless~\cite{Chen:1999tn}
and $1 / M^{7/2} Q^{1/2}$ in pionful~\cite{Valderrama:2014vra}.
The ${\rm LO}$ is set by the one-body charge current
\begin{eqnarray}
\langle \vec{p}\,' | J^0(\vec{q}) | \vec{p} \rangle &=& e\,
\delta (\vec{p}' - \vec{p} - \frac{\vec{q}}{2}) \, , 
\end{eqnarray}
that scales as $Q^{-3}$.
Therefore $D(\Lambda)$ and $Q(\Lambda)$ enter at ${\rm N^5LO}$
in NDA~\cite{Phillips:2003jz},
${\rm N^3LO}$ in pionless~\cite{Chen:1999tn}
and ${\rm N^{9/2}LO}$ in pionful~\cite{Valderrama:2014vra}.

The magnetic form factor describes the response of
the deuteron to a transversal current.
It is analogous to the matrix elements that appear in the deuteron breakup
reactions. The power counting of the contact currents is identical to that case
except for the difference that the initial and final states are in the
$^3S_1-{}^3D_1$ channel.
In NDA~\cite{Phillips:2003jz} the contact contribution
to $G_M$ enters at ${\rm N^3LO}$,
in pionless~\cite{Chen:1999tn} at ${\rm NLO}$
and in pionful~\cite{Valderrama:2014vra} at ${\rm N^{5/2}LO}$.

To close the discussion we will briefly consider proton-proton fusion
($pp \to d\,e^{+}\,\nu_e$), which is a weak process.
The average momenta of protons at the core of the sun is about $1\,\rm keV$.
From the perspective of nuclear physics these are extremely low momenta and
we expect solar proton-proton fusion to happen almost entirely 
via the s-wave transition $^1S_0 \to ^3S_1$.
For this reaction the relevant one-body weak current is axial
and takes the form:
\begin{eqnarray}
\langle \vec{p}\,' | \vec{A}_{1B}(\vec{q}_{e^+} + \vec{q}_{\nu_e}) |
\vec{p} \rangle &=&
- G_V g_A \left[ \vec{\sigma}_1 \tau_1^{-} + \vec{\sigma}_2 \tau_2^{-}
\right] \nonumber \\ &\times&
\delta^{(3)}(\vec{p}\,' - \vec{p} - \frac{1}{2}\vec{q}_{e^+} -
\frac{1}{2}\vec{q}_{\nu_e}) \, ,
\end{eqnarray}
where $G_V$ is the weak vector coupling, $g_A$ the axial-to-vector ratio,
$\vec{q}_{e^{+}}$ and $\vec{q}_{\nu_e}$ are the momenta of the final positron
and neutrino, $\sigma_i$ the spin of the nucleon $i = 1,2$ and
$\tau^{-}_i$ an isospin operator that turns a proton into a neutron, i.e.
$\tau^{-} | p \rangle = | n \rangle$.
This one-body current scales as $Q^{-3}$ owing to the delta.
Meanwhile the simplest axial two-body current
that we can construct takes the form
\begin{eqnarray}
\langle \vec{p}\,' | \vec{A}_{1B} | \vec{p} \rangle = A(\Lambda)\,\vec{\beta}
\, ,
\end{eqnarray}
with $\beta$ a pseudovector containing the spin and isospin components
(the exact form of this current can be checked in Ref.~\refcite{Park:2002yp}).
In NDA this two-body current scales as $Q^{0}$,
i.e. $\rm N^3LO$ relative to the one-body current.
In pionless and pionful the $A(\Lambda)$ coupling will be enhanced
by a factor $(M / Q)^2$ and $(M / Q)^{5/4}$ respectively,
that is, $\rm NLO$ and $\rm N^{7/4}LO$.

\subsection{The Three Body Contact in Pionless}

Other application is the power counting of the three-body contact-range
interaction in pionless EFT~\cite{Bedaque:1998kg,Bedaque:1998km,Bedaque:1999ve,Griesshammer:2005ga}.
A three-body system of identical bosons that interacts via two-body
contact-range interactions can bind, more so if the two-body system
contains a bound or virtual state.
The wave function can be expressed as a sum of three components
(the Fadeev components)
\begin{eqnarray}
| \Psi_{3B} \rangle = 
\langle \vec{p}_1 \, \vec{k}_{23} | \psi_{3B} \rangle +
\langle \vec{p}_2 \, \vec{k}_{31} | \psi_{3B} \rangle +
\langle \vec{p}_3 \, \vec{k}_{12} | \psi_{3B} \rangle \, ,
\end{eqnarray}
where all the components are identical: we have bosons and the wave function
is invariant under permutation of the particle labels.
The momenta $\vec{p}_i$, $\vec{k}_{ij}$ are the Jacobi momenta, which are
\begin{eqnarray}
\vec{p}_1 &=& \frac{2}{3}\,\vec{q}_1 - \frac{1}{2}\,(
\vec{q}_2 + \vec{q}_3 ) \, , \\
\vec{k}_{23} &=& \frac{1}{2}\,( \vec{q}_2 - \vec{q}_3 ) \, ,
\end{eqnarray}
plus permutations, with $\vec{q}_{1,2,3}$ the momenta of each of the particles.
For the case of s-wave, non-perturbative contact-range forces,
there is a compact ansatz for the Fadeev component $\psi_{3B}$
\begin{eqnarray}
\langle \vec{p} \, \vec{k} | \psi_{3B} \rangle =
\frac{a(p)}
{\frac{3}{4}\,p^2 + k^2 + \gamma_3^2}
\end{eqnarray}
where $\gamma$ is the wave number of the three-body system and
$a(p)$ is a function of the Jacobi momentum $p$.
If the mass of each of the identical particles is $M$,
the binding energy is $E_B = - \gamma_3^2 / M$.
The function $a(p)$ is given by
\begin{eqnarray}
a(p) \propto \frac{1}{p^2}\,f(p) \, ,
\end{eqnarray}
at large momenta, where $f(p)$ is an oscillatory function of the type
\begin{eqnarray}
f(p) = \sin{ \left[ s_0 \log{\frac{p}{p_0}} \right]} \, ,
\end{eqnarray}
where $s_0 \simeq 1.0064$ and $p_0$ is a reference momentum
that we will discuss in a moment. 

The details of how one reaches the three-body wave-function are irrelevant
for our purposes here, but can be consulted
in the literature~\cite{Bedaque:1998kg,Bedaque:1998km}.
The relevant point here is that we can make interesting conclusions
about the power counting from inspecting the three-body bound-state
wave function.
The value of the reference momentum $p_0$ cannot be determined from
the two-body contact interactions unambiguously.
If we include a cut-off $\Lambda$ then we can calculate a $p_0(\Lambda)$,
but it does not converge to a specific value as we increase $\Lambda$.
That is, the value of $p_0$ depends on the short-range physics.
But since the two-body short-range physics are already included in the EFT,
the conclusion is that there is a contact-range three-body force
also entering at ${\rm LO}$.

Now we rederive this result with the calculation of
the anomalous dimension of the three-body contact-range coupling.
In momentum space the lowest order three-body contact-range potential reads
\begin{eqnarray}
\langle \vec{p}\,' , \vec{k}\,' | V_{C} | \vec{p} , \vec{k} \rangle = C_3 \, ,
\end{eqnarray}
where the naive estimation of its size is $C_3 \sim 1/M^3$.
The matrix element of this potential when sandwiched between the $\Psi_{3B}$
wave function is
\begin{eqnarray}
\langle \Psi_{3B} | C_3 | \Psi_{3B} \rangle
\propto 
\langle \psi_{3B} | C_3 | \psi_{3B} \rangle = C_3\,
{\left[ \int_{\Lambda} \frac{d^3\vec{p}}{(2\pi)^3} \frac{d^3\vec{k}}{(2\pi)^3} 
\frac{a(p)}{\frac{3}{4} p^2 + k^2 + \gamma^2} \right]}^2 \, ,
\end{eqnarray}
which diverges as
\begin{eqnarray}
\langle \Psi_{3B} | C_3 | \Psi_{3B} \rangle
\propto C_3\,\Lambda^4 \, .
\end{eqnarray}
We end up with the RGE
\begin{eqnarray}
\frac{d}{d\Lambda}\left[ C_3(\Lambda)\, \Lambda^4 + \dots \right] = 0 \, ,
\end{eqnarray}
which implies a $(M / Q)^4$ enhancement over the NDA estimation.
The non-relativistic three-body propagator counts as $Q^4$:
\begin{eqnarray}
I_3(E) &=& \int \frac{d^3\vec{p}}{(2\pi)^3} \frac{d^3\vec{k}}{(2\pi)^3} G_0(E)
\nonumber \\
&=& 
\int \frac{d^3\vec{p}}{(2\pi)^3} \frac{d^3\vec{k}}{(2\pi)^3}
\frac{1}{M E - M (\frac{3}{4} p^2 + k^2)} \sim Q^4 \, ,
\end{eqnarray}
which at the end means that $C_3$ has to be iterated to all orders
because $C_3 \sim C_3 I_3(E) C_3$.
That is, $C_3$ enters at ${\rm LO}$.

We can easily apply these ideas to the three-nucleon system,
though there are a few complications owing to spin, isospin and
the fact that nucleons are fermions.
The presence of spin and isospin degrees of freedom allows the spatial part
of the wave function to be partially or fully symmetric
for specific configurations.
When that happens the conclusions that we derived for three-boson system
may apply to the three-nucleon system as well~\cite{Bedaque:1999ve}.
This is the case for the triton, where the three-nucleon contact enters
at ${\rm LO}$ in the pionless EFT.
This also happens for neutron-deuteron scattering in the spin-$1/2$
configurations (the doublet), for the simple reason that this is
the same channel as the triton.
Trivially this conclusion applies too for the $^3{\rm He}$ nucleus and doublet
proton-deuteron scattering at ${\rm LO}$, though here we have the additional
complication of Coulomb~\cite{Ando:2010wq}.
Recently it has been discovered that Coulomb is able to modify 
the counting at ${\rm NLO}$~\cite{Vanasse:2014kxa,Konig:2014ufa},
but only if it is treated non-perturbatively~\cite{Konig:2015aka}.
On the contrary for the spin-$3/2$ configurations (the quartet)
of the three-nucleon system the size of the three-nucleon contact
is the one expected in NDA.
The reason is that the spatial wave function cannot be symmetric
for the quartet. If we visualize the quartet as a nucleon scattering off
a deuteron it is clear that this nucleon must be in a P-wave
with respect to the deuteron.
The wave function has a certain resemblance with the one we have studied
for the three-boson system.
However the piece that depends on $\vec{p}$ is not of the type $a(p)$
but rather $\vec{\beta} \cdot \vec{p} \, a(p)$. In addition
the power-law dependence of $a(p)$ is much more suppressed
at high momenta, a change that modifies the anomalous dimension
of the three-body coupling.
The outcome is that the contact coupling is not enhanced
(and might even be demoted~\footnote{Here we will not consider
the possibility of the demotion of a coupling.
The reason is that even if the anomalous dimension of a coupling is positive,
this only induces small corrections over the $1/M^d$ baseline
value of the coupling that is used as a boundary condition
at $\Lambda = M$.}) in this case~\cite{Griesshammer:2005ga}.

The previous ideas can be extended to triton and $^3{\rm He}$ reactions.
As we already explained, in a reaction the few-nucleon part of
the wave function factors out when computing the matrix elements.
The conclusion is that contact few-body operators involving external probes
have the same type of enhancements as contact few-body forces.
On practical terms a three-body operator involving the triton (or the doublet)
in the initial and final states will be enhanced by the factor
${\left( M / Q \right)}^4$.
If one of the initial and final channels is the triton/doublet and the other
is the quartet, the enhancement will be ${\left( M / Q \right)}^2$.
Finally a reaction only involving an initial and final quartet
is not expected to be enhanced.

\section{Conclusions}

In this manuscript I have attempted to illustrate the application of
Wilsonian renormalization to nuclear EFT.
The starting point is the non-relativistic two-body scattering problem.
The requirement that the scattering amplitude is invariant under changes
in the cut-off generates RGEs for the couplings of the EFT. 
The RGEs can be calculated easily from the two-body wave functions.
The solution of the RGEs in the infrared --- as we change the momentum
cut-off from $M$ to $Q$ --- determines the power counting of the EFT.
There are different power countings depending on the initial assumptions
about scattering at low energies.
In agreement with the previous literature,
for non-relativistic two-body systems that interact via a regular potential
there are two general power countings, the natural and
the unnatural one~\cite{Birse:1998dk,Barford:2002je}.
If the potential is non-perturbative and singular, as happens
in nucleon-nucleon scattering in the triplet partial waves,
there is a unique power counting and the idea of fine tuning
is not that crucial: all the scattering lengths are equally
fine-tuned.
Yet this conclusion is not universally agreed upon
within the context of RGA~\cite{Birse:2005um}.

There is also the interesting observation that the solution of the RGEs
for a particular coupling is connected to its anomalous
dimension~\cite{Valderrama:2014vra}.
The anomalous dimension refers to how a coupling changes
under a rescaling of the cut-off.
It turns out that the calculation of the anomalous dimension is trivial,
merely involving the evaluation of a matrix element between
the EFT wave functions.
Incidentally this development also makes it easy to determine
power counting beyond the two-body system. 
We have also shown the equivalence among Wilsonian renormalization,
standard or ultraviolet renormalization and the analysis of
the residual cut-off dependence as methods
to uncover the power counting.
But there are also open problems regarding the renormalization of singular
interactions.
For attractive singular potentials the previous equivalence is not
completely proven yet (which explains the existence of slightly different
versions of the power counting in the literature~\cite{Birse:2005um,Valderrama:2009ei,Valderrama:2011mv,Long:2011qx,Long:2011xw}),
though the evidence pointing towards this direction is convincing.
In the case of a repulsive singular potential the big unsolved issue is
whether its power counting is the same as for the attractive case
or if it follows NDA.
This question has so far evaded a satisfactory analysis.

The advantage of the anomalous dimension is that it can be easily calculated
in the case of reactions of external electroweak probes acting
on the two-nucleon system.
The power counting is essentially the one of the two-nucleon system
for the simple reason that the probes we are considering
can be described with a plane wave and factored out of the RGEs.
As a consequence this idea does not hold if the external probe
is a third nucleon. 
For the three-nucleon system we must calculate the wave functions again
if we want to be able to determine the power counting.
The behaviour of the three-body wave functions is well-known
for a contact-range potential, from which we can independently
reproduce the power counting of pionless EFT for the triton
and neutron-deuteron scattering.
In the future once the power-law behaviour of the wave functions of
the triton is properly analyzed it will be possible to determine
the power counting of the three-nucleon system in pionful EFT
(though there are preliminary results~\cite{Birse:2009my}).

\section*{Acknowledgements}

I am very grateful to D.R. Phillips and E. Ruiz Arriola
for the collaborations in which this review is largely based.
I also thank D.R. Phillips for a careful reading of this manuscript.
I would also like to thank M.C. Birse, B. Long, M. S\'anchez S\'anchez and
U. van Kolck for discussions that have influenced and shaped
many of the issues dealt with in this manuscript.

\appendix

\section{The Delta-Shell Potential}
\label{app:delta}

In this appendix we derive the solution of the Schr\"odinger equation
for a delta-shell potential.
The starting point is
\begin{eqnarray}
- u_k'' + 2\mu\,\left[ V(r; r_c) + V_C(r; r_c) \right] u_k = k^2 u_k  \, ,
\end{eqnarray}
where $u_k$ is the reduced wave function, $k$ the momentum,
$V$ the finite-range potential and $V_C$ the delta-shell potential.
The form of $V_C$ is
\begin{eqnarray}
V_C(r; r_c) = \frac{C_k(r_c)}{4\pi r_c^2}\,\delta(r - r_c) \, ,
\end{eqnarray}
where $C_k(r_c)$ can be expanded in powers of $k^2$.
We can see that $V_C$ only acts at $r = r_c$.
Now we integrate the Schr\"odinger equation in the vicinity of $r_c$
\begin{eqnarray}
\int_{r_c - \epsilon}^{r_c + \epsilon}\,
\left( 
-u_k'' + 2\mu\,\left[ V(r; r_c) + V_C(r; r_c) \right] u_k 
\right) \,dr =
k^2 \int_{r_c - \epsilon}^{r_c + \epsilon} u_k (r)\, dr  \, ,
\end{eqnarray}
with $\epsilon$ a small positive number.
The evaluation of the following pieces is direct
\begin{eqnarray}
\int_{r_c - \epsilon}^{r_c + \epsilon} u_k'' (r)\, dr  &=& 
u_k'(r_c + \epsilon) - u_k'(r_c - \epsilon) \, , \\
2\mu\,\int_{r_c - \epsilon}^{r_c + \epsilon}\,V_C(r; r_c) u_k(r)\,dr &=& 
2\mu\,\frac{C_{k}(r_c)}{4 \pi r_c^2}\,u_k(r_c) \, ,
\end{eqnarray}
while the two remaining pieces vanish in the $\epsilon \to 0$ limit
\begin{eqnarray}
\lim_{\epsilon \to 0} \,
\int_{r_c - \epsilon}^{r_c + \epsilon}\,2\mu\,V(r; r_c) u_k(r)\,dr &=& 0 \, , \\
\lim_{\epsilon \to 0} \, 
k^2 \int_{r_c - \epsilon}^{r_c + \epsilon} u_k (r)\, dr  &\to& 0 \, ,
\end{eqnarray}
the reason being that the integrand is bounded in the region around $r_c$.
We can see that while $u_k$ is continuous at $r = r_c$,
$u_k'$ develops a discontinuity.
Putting the pieces together for $\epsilon \to 0$ we arrive at
\begin{eqnarray}
\frac{u_k'(r_c^+)}{u_k(r_c^+)} - \frac{u_k'(r_c^-)}{u_k(r_c^-)} 
=  2\mu\,\frac{C_{k}(r_c)}{4 \pi r_c^2}\,u_k(r_c) \, .
\end{eqnarray}
Finally, expanding $C_k$ in powers of $k^2$ we obtain Eq.~(\ref{eq:RGE}).


\end{document}